# Spin injection in Si-based ferromagnetic tunnel junctions with MgO/MgAl$_2$O$_4$ barriers: Experimental and theoretical investigation of barrier thickness-dependent spin tunneling efficiency

Baisen Yu[1], Shoichi Sato[1,2], Masaaki Tanaka[1,2,3], and Ryosho Nakane[1,3,4]

[1]*Deptartment of Electrical Engineering and Information Systems, The University of Tokyo, 7-3-1 Hongo, Bunkyo-ku, Tokyo 113-8656, Japan*

[2]*Center for Spintronics Research Network (CSRN), The University of Tokyo, 7-3-1 Hongo, Bunkyo-ku, Tokyo 113-8656, Japan*

[3]*Institute for Nano Quantum Information Electronics, The University of Tokyo, 4-6-1 Meguro-ku, Tokyo 153-8505, Japan*

[4]*Systems Design Lab (d.lab), The University of Tokyo, 7-3-1 Hongo, Bunkyo-ku, Tokyo 113-8656, Japan*

**Abstract**

We have experimentally and theoretically investigated the spin transport in Fe/Mg/MgO/MgAl$_2$O$_4$/$n^+$-Si(001) ferromagnetic tunnel junctions on a Si substrate, by systematically varying the thickness combination of amorphous MgO and MgAl$_2$O$_4$ tunnel barrier layers with a sliding shutter between the evaporation sources and substrate during electron-beam evaporation. A technical advantage of MgAl$_2$O$_4$ is that a continuous and flat thin film is realized on a Si substrate even when the MgAl$_2$O$_4$ thickness is as thin as 0.5 nm, unlike MgO, which enables us to examine the spin transport in a thinner range of the tunnel barrier thickness. Our distinct finding is as follows: When the Fe/Mg/MgO interface is used on the top side, the spin polarization $P_S$ of tunneling electrons increases at 10 K as the total MgO/MgAl$_2$O$_4$ tunnel barrier thickness ($t_{ox}$ = 0.47 − 1.4 nm) is increased, regardless of different thickness combinations, and $P_S$ shows saturation-like behavior when $t_{ox}$ is above 1.1 nm. Since this feature cannot be explained by the well-known conductivity mismatch in semiconductor-based ferromagnetic tunnel junctions, we propose a simple phenomenological tunneling model based on two different direct tunneling paths, which have higher/lower spin polarizations with longer/shorter decay lengths. Our numerical calculation reproduces the relationship between the spin polarization $P_S$ and total tunnel barrier thickness $t_{ox}$ in the experiments, indicating that the dominant mechanism is an increasing contribution of the lower spin polarization path as $t_{ox}$ is decreased. We discuss possible origins for this phenomenon including intrinsic and extrinsic tunneling mechanisms. Our analysis method provides an insight into the detailed spin transport physics in semiconductor-based ferromagnetic junctions, particularly, with a very thin tunnel barrier layer.

# I. Introduction

Si-based spin transport devices are promising candidates for next-generation mobile electronics with low power consumption due to their nonvolatile and reconfigurable spin functionalities [1−6]. Among them, Si-based spin metal-oxide-semiconductor field-effect transistors (spin MOSFETs) are particularly attractive [7−13], since their nonvolatility offers a potential for direct integration with conventional complementary MOS (CMOS) technology on a Si platform, enabling a wide range of energy-efficient circuits applications. A spin MOSFET shares the same basic structure as an ordinary MOSFET, except that its source (S) and drain (D) electrodes are replaced with ferromagnetic materials or ferromagnetic tunnel junctions. In this manner, the spin-dependent device resistance, i.e., magnetoresistance (MR), can be changed by the relative magnetization configuration between the S and D electrodes. For practical applications, a high MR ratio, defined by the normalized change in MR, is required. The MR is governed by three fundamental processes: the injection of spin-polarized electrons (spin injection) from the ferromagnetic S electrode to the Si channel at the S junction, electron conduction via the two-dimensional (2D) channel with no spin flip, and the detection of spin-polarized electrons (spin detection) at the D junction. In our previous work [12−14], we progressively revealed the physical mechanism of the spin transport through Si 2D channels in spin MOSFETs with S/D ferromagnetic Fe/MgO/Si tunnel junctions and a $SiO_2$/Si gate stack. We demonstrated that the effective spin diffusion length in the Si 2D channel reaches several tens of micrometers at room temperature, facilitated by the spin drift effect [13]. These results highlight the strong potential of Si for practical spintronic applications. The remaining issue is the low MR ratio (<1%), which is mainly attributed to the high tunnel junction resistance [13].

In parallel, extensive research has focused on the spin transport in ferromagnetic metal (FM)/insulator (I)/$n^+$-Si tunnel junctions and their associated material properties [15−25]. Through a detailed analysis of the bias-dependent features of spin signals for a Fe/MgO/$SiO_x$/Si tunnel junction, we clarified the physical mechanism of spin injection and detection using an original band diagram model for Fe/MgO/Si tunnel junctions [19]. We found a distinct phenomenon, termed "spin accumulation saturation (SAS)", under sufficiently high bias conditions. This feature suggests that ferromagnetic tunnel junctions with a moderate resistance are essential to maintain high spin polarization $P_S$ while a large current is realized [19]. Our previous studies further revealed that the formation of a magnetically dead layer at the FM/I interface critically degrades the amplitude of a spin signal. This issue can be effectively resolved by inserting an ultrathin Mg layer at the interface [15,16]. In addition, we demonstrated that the spin injection/detection efficiencies are significantly improved by reducing the defect density at the MgO/Si interface through the insertion of a thin $SiO_x$ interlayer [18]. Regarding the selection of insulating barriers, despite efforts to explore alternative materials, such as $Al_2O_3$, $SiN_xO_y$ [17,20], efficient spin injection on a Si platform has been predominantly achieved using MgO-based tunnel junctions. In particular, $P_S$ exceeding 20% at 4 K have been reported only in MgO-based tunnel junctions [16,18,21,23]. It is noteworthy that the well-known conductivity mismatch problem cannot account for low $P_S$ observed in the junctions based on other barriers, since all the reported structures exhibit barrier resistances significantly higher than the spin resistance of $n^+$-Si [17,20,22]. These results highlight the need to understand why MgO-based junctions uniquely enable efficient spin injection, from both material science and spin transport perspectives.

On the other hand, spin injection in ferromagnetic tunnel junctions with a very thin barrier thickness (< 1 nm) has been rarely studied, which may be due to difficulties in fabricating uniform and smooth insulating thin films on Si. For example, MgO tends to exhibit poor wetting behavior and surface roughness when directly deposited on Si, making it difficult to form a high-quality ultrathin barrier [16]. As a result, most spin injection experiments were conducted using junctions with relatively thick tunnel barriers (>1 nm), which result in high junction resistance-area ($RA$) products and low MR ratios [13,16,18]. Since reducing the barrier thickness is one of the most effective strategies to lower $RA$ values for higher MR ratios, it is essential to understand how $P_S$ varies with the barrier thickness (or $RA$). A review of recent studies reveals a consistent trend: junctions that exhibit high $P_S$ values tend to have higher $RA$ values [16,18,23]. For instance, a $P_S$ of 41% was achieved at 4 K in a Fe/Mg/MgO/SiO$_x$/$n^+$-Si junction with a $RA$ of ~120,000 Ω·μm$^2$, whereas a reduced $P_S$ of 12% was observed at 4 K in a Fe/Mg/MgO/$n^+$-Si junction with a $RA$ of ~3800 Ω·μm$^2$. These facts indicate a physical correlation between $P_S$ and $RA$ of FM/I/Si junctions, but it has not been comprehensively studied due to variations in FM and I materials, as well as different fabrication methods, in those junctions. Therefore, to clarify the physical correlation and gain a more comprehensive understanding of spin transport mechanisms, it is necessary to investigate the spin injection through FM/I/Si junctions fabricated on a Si substrate with systematically varied barrier thicknesses while the identical material composition and fabrication method are used.

The purpose of this study is to experimentally and theoretically investigate the relationship between $P_S$ and barrier thickness using a Fe/MgO/MgAl$_2$O$_4$/$n^+$-Si structure fabricated on a $n^+$-Si substrate while various combinations of MgO and MgAl$_2$O$_4$ thicknesses are utilized. First, we examine the material properties of Fe/MgO/MgAl$_2$O$_4$/$n^+$-Si junctions and demonstrate that the insertion of a MgAl$_2$O$_4$ layer between MgO and Si is essential for significant spin signals across both thin and thick barrier ranges. Then, we study the electron and spin transport properties of these junctions. The estimated $P_S$ values exhibit a distinctive dependence on the total barrier thickness, which shows a degree of generality when compared with previous studies. To interpret this thickness dependence, we propose a phenomenological "two-path" model that incorporates two tunneling paths with different spin polarizations and tunneling decay characteristics. A good agreement between experimental results and theoretical predictions validates the effectiveness of our model, and physical origins of our phenomenological model are discussed.

## II. Sample preparation and characterization

### II-A. Sample preparation and fabrication

Figure 1(a) shows all the layered structures examined in this study, where their names (type I−VII) will be used throughout this study. The preparation method of the type I structure is similar to that in our previous work [17,18]. First, a phosphorus-doped $n^+$-Si(001) (1×10$^{20}$ cm$^{-3}$) substrate was thermally oxidized to form a 110-nm-thick SiO$_2$ layer that was used to isolate each tunnel junction. Then, partial areas of the oxidized substrate were etched by HF in two steps to obtain a clean and flat H-terminated Si surface with circular shapes and areas $A$ = 25, 250, 2500, and 25000 μm$^2$. Second, the substrate was loaded into an ultra-high vacuum (UHV) chamber with a based pressure of 1×10$^{−7}$ Pa, and thermally cleaned at a substrate temperature $T_S$ = 700 °C for 10 min. Subsequently, at $T_S$ = 650 °C, a MgAl$_2$O$_4$ layer was deposited by electron beam evaporation of a single-crystal MgAl$_2$O$_4$ source. Next, after $T_S$ was cooled down to room temperature (25 °C), a MgO layer was deposited by electron beam evaporation of a single-crystal MgO source. During the depositions, a shutter was used to change the MgO thickness ($t_{MgO}$) and MgAl$_2$O$_4$ thickness ($t_{MAO}$), respectively. In this manner, a MgO/MgAl$_2$O$_4$ bilayer barrier structure can be fabricated with several $t_{MgO}$ and $t_{MAO}$ combinations. Here, we define the total oxide thickness $t_{ox} = t_{MAO} + t_{MgO}$. The layer thicknesses and deposition rates of the MgAl$_2$O$_4$ and MgO layers were measured *in-situ* by a quartz crystal microbalance (QCM) sensor and calibrated by cross-sectional transmission electron microscopy (TEM) images. After the MgAl$_2$O$_4$ and MgO depositions, the substrate was transferred into a molecular beam epitaxy (MBE) chamber without breaking UHV. Then, Al(15 nm)/Mg(1 nm)/Fe(4 nm)/Mg(0.5 nm) multilayer was deposited at room temperature using Knudsen cells [16]. To prevent the multilayer from oxidization, the substrate was transferred into a sputtering chamber without breaking UHV, where a 10-nm Pt layer was deposited. Third, immediately after being exposed to air, a 150-nm-thick Al layer was deposited on the substrate by thermal evaporation, and top electrode was fabricated for multiple junctions using UV lithography followed by Ar-milling. Finally, an Al layer of 150 nm was deposited on the back of the substrate just after removing native oxide by HF etching. The fabrication methods for type II−VII structures are essentially the same as that for a type I structure, except that a SiO$_2$ layer was not formed prior to the deposition and the back-side Al electrode was omitted. For type II−IV structures, after the depositions of MgAl$_2$O$_4$ and MgO layers, the substrates were then directly removed from the UHV chamber for surface roughness evaluation using atomic force microscopy (AFM). For type V−VII structures, the Al/Mg/Fe/Mg multilayers were deposited on MgO($t_{MgO}$)/MgAl$_2$O$_4$($t_{MAO}$) to study the influence of the oxide layers on the magnetic properties of the Fe layer, characterized by superconducting quantum interference device (SQUID) magnetometry.

### II-B. Surface roughness estimation of MgO/$n^+$-Si, MgAl$_2$O$_4$/$n^+$-Si, and MgO/MgAl$_2$O$_4$/$n^+$-Si structures

Fabrication of a thin tunnel barrier layer with excellent surface flatness on Si is one of the essential requirements for spin injection [16]. To investigate whether the introduction of a MgAl$_2$O$_4$ layer on Si(001) can result in improved surface flatness compared with a single MgO barrier or not, surface morphologies of the type II−IV structures were studied using AFM images. Figures 2(a) and (b) show the surface morphologies of type II and type III structures with $t_{MgO}$ = 0.5 nm and $t_{MAO}$ = 0.5 nm, respectively. Whereas both structures exhibit continuous surface, the MgAl$_2$O$_4$ layer has improved flatness compared with the MgO layer, as indicated by the fewer black dark regions corresponding to large peak-to-valley differences. This observation is quantitatively supported by the root mean square (RMS) values of the surface; 0.336 nm for the MgO layer (type II) in Fig. 2(a) and 0.193 nm for the MgAl$_2$O$_4$ layer (type III) in Fig. 2(b). On the other hand, Figs. 2(c) and (d) show surface morphologies of a type IV structure with the MgO/MgAl$_2$O$_4$ bilayer combinations of $t_{MgO}$ = 3.3 Å/$t_{MAO}$ = 1.4 Å and $t_{MgO}$ = 5.5 Å/$t_{MAO}$ = 4.1 Å, respectively. The corresponding RMS values were estimated to be 0.156 nm and 0.145 nm, respectively. It is found that the MgO/MgAl$_2$O$_4$ bilayer (type IV) exhibits superior flatness compared with the single MgO (type II) and single MgAl$_2$O$_4$ (type III) structures. Moreover, the RMS values estimated in Figs. 2(c) and (d) are almost unchanged. This confirms that the surface roughness of the MgO/MgAl$_2$O$_4$ bilayer is nearly independent of the layer thickness. Thus, we concluded that the MgAl$_2$O$_4$ thin layer acts as an effective buffer layer for the formation of a smooth MgO film, and that the MgO/MgAl$_2$O$_4$ bilayer is very promising to be used as a tunnel barrier for efficient spin injection.

### II-C. Characterization of Fe/Mg/MgO/MgAl$_2$O$_4$/$n^+$-Si by transmission electron microscopy (TEM)

Figures 3(a) and (b) show cross-sectional transmission electron microscopy (TEM) images of a type I structure with MgO/MgAl$_2$O$_4$ combinations of $t_{MgO}$ = 3.3 Å/$t_{MAO}$ = 7.6 Å and $t_{MgO}$ = 9.9 Å/$t_{MAO}$ = 7.6 Å, respectively, where the incident electron beam was aligned along the [110] crystallographic axis of Si. The three colored layers, from bottom to top, were identified as the Si substrate (dark), MgO/MgAl$_2$O$_4$ bilayer (light), and Fe/Mg layer (dark). Based on the TEM features, the Fe layer and MgO/MgAl$_2$O$_4$ bilayer were found to be polycrystalline and amorphous, respectively. The total oxide thickness was estimated to be $t_{ox}$ = 1.15 ± 0.10 nm in Fig. 3(a) and $t_{ox}$ = 1.88 ± 0.13 nm in Fig. 3(b). These values were used as the reference to calibrate the depositing rates of MgAl$_2$O$_4$ and MgO estimated by the QCM sensor. Although MgO and MgAl$_2$O$_4$ cannot be distinguished within the light amorphous layers due to their similar chemical compositions and thin thicknesses, the increase of $t_{MgO}$ from 3.3 Å to 9.9 Å under a fixed $t_{MAO}$ = 7.6 Å is clearly reflected by the change in $t_{ox}$. The MgAl$_2$O$_4$/Si interface appears atomically sharp with minimal interfacial fluctuation, whereas the interface between Fe/Mg and MgO exhibits surface roughness of approximately 0.1 nm. These roughness values estimated by the TEM images are consistent with the RMS values estimated from the AFM images shown in Figs. 2(c) and (d).

### II-D. Characterization of layered structures by SQUID: magnetic properties

To study the influence of various oxide layers on the magnetic properties of the Fe thin film, as shown in Figure 4, we measured the magnetization *vs.* in-plane magnetic field ($M$–$H_{//}$) curves at 10 K for type V, type VI, and type

VII structures, corresponding to $t_{MgO}$ = 0.5 nm, $t_{MAO}$ = 0.5 nm, and $t_{MAO}$ = 0.5 nm/$t_{MgO}$ = 0.5 nm, respectively. The gray, blue, and pink curves represent type V, type VII, and type VI structures, respectively. The saturation magnetization $M_S$ and remanent magnetization $M_r$ are defined as the $M$ values at $H_{//}$ = 300 and 0 Oe, respectively. The coercive field $H_C$ is defined as the average of the $|H_{//}|$ values at which $M$ = 0. Type V (gray) and type VII (blue) structures exhibit similar values of $M_S \sim$ 1550 emu/cm$^{-3}$ and $H_C \sim$ 10 Oe, while a slight difference is observed in the remanence ratio (= $M_r/M_S$): type V and type VII have $M_r/M_S$ = 1 and 0.77, respectively. On the other hand, the $M$–$H_{//}$ curve of the type VI structure (pink) shows distinct features: a reduced $M_S$ = 1250 emu/cm$^{-3}$, an enhanced $H_C$ = 36 Oe, and a comparable $M_r/M_S$ value of 0.95. These clearly reveal that the magnetic properties of the Fe thin layer are sensitive to the underlying oxide layer. In particular, the degraded $M_S$ value in the Fe/Mg/MgAl$_2$O$_4$ structure (type VI) can be related with the spin injection efficiency.

## III. Measurement methods of spin Hanle signals and *I-V* characteristics for estimating spin polarization $P_S$ and resistance-area product *RA*

Figure 5 shows a schematic illustration of the measurement setup for three-terminal Hanle (3TH) signals and *I-V* characteristics. In the 3T Hanle measurement, the voltage change $\Delta V_{3T}$ is measured with a constant current $I_B$ between the target junction and backside of the Si substrate, while a magnetic field $H$ is swept between ±3 kOe. When we apply an in-plane magnetic field $H_{//}$ parallel to the Si(001) plane, an inverted 3TH (I-3TH) signal $\Delta V_{3T}^I$ is obtained. Although an I-3TH signal is not related to *true* spin injection, it can be used to qualitatively analyze a magnetically dead layer forming at a ferromagnetic metal/oxide interface [15]. When we apply a perpendicular magnetic field $H_\perp$ along the Si [001] direction, we observe a 3TH signal consisting of a narrower 3TH (N-3TH) signal $\Delta V_{3T}^N$ and a broader 3TH (B-3TH) signal $\Delta V_{3T}^B$. Their typical half widths at half maximum (HWHM) are a few-tens and a few-hundreds Oe, respectively [15,16]. In this study, we mainly focus on the analysis of N-3TH signals, since $\Delta V_{3T}^B$ is not related to the *true* spin injection. The detailed estimation procedures of $\Delta V_{3T}^N$ are similar to those presented in our previous work [18]. When $\Delta V_{3T}^N$ is obtained, the target junction is applicable to spin injectors and detectors. In the *I-V* measurements, the bias current $I_B$ is measured as a function of the junction voltage drop $V_B$. The junction resistance ($R_J$) and resistance-area product (*RA*) of a tunnel junction are defined by $V_B / I_B$ and $A \times V_B / I_B$, respectively, which are functions of $I_B$ (or $V_B$). The positive (negative) polarity of $I_B$ corresponds to the spin injection (extraction) geometry, where electrons are injected from the Fe layer to the Si substrate (electrons are extracted from the Si substrate to the Fe layer). In this study, we mainly use negative $I_B$, because a significant $\Delta V_{3T}^N$ signal cannot be observed in the spin injection geometry ($I_B > 0$), at which spin detection efficiency is nearly zero due to the strong nonlinear junction properties [19]. The spin polarization $P_S$ and spin lifetime $\tau_s$ are estimated through the fitting with the following formula [16]

$$\Delta V_{3T}^N(H_\perp) = \Delta V_0 \sqrt{\frac{1+\sqrt{1+(\gamma H_\perp \tau_S)^2}}{2+2(\gamma H_\perp \tau_S)^2}}, \quad (1)$$

$$\Delta V_0 = P_{inj} P_{det} J_B \rho \lambda_{sf} = P_S^2 J_B \rho \lambda_{sf},$$

where $J_B$ is the current density defined by $I_B/A$, $\gamma$ is the gyromagnetic ratio, $\lambda_{sf}$ = 1 μm is the spin diffusion length in $n^+$-Si [16], $\rho$ = 1 mΩ·cm is the resistivity of $n^+$-Si [16], and $P_{inj}$ and $P_{det}$ are the spin injection and detection polarizations, respectively. For each fitting and estimation, the error margins are calculated based on root-mean-square deviation (RMSD) method [see Sec. S1 in S.M[29]]. In our previous work, we theoretically proved that $P_{inj}$ and $P_{det}$ are correlated in a nearly linear manner [19]. Thus, $P_S$ is a key factor to evaluate the changing trends of both $P_{inj}$ and $P_{det}$ as junction structures are changed.

# IV. Spin injection through type I structures: ferromagnetic tunnel junctions based on MgAl$_2$O$_4$ single layers and MgO/MgAl$_2$O$_4$ bilayers

The atomically flat MgO/MgAl$_2$O$_4$ and MgAl$_2$O$_4$ thin layers, characterized by the AFM and TEM images in Sec. II, indicate that they are promising for the use as insulating (I) tunnel barriers in ferromagnetic metal (FM)/I/Si structures for spin injection. Highly efficient spin injection does not only depend on the flatness of the I layer, but also on interfacial conditions at the FM/I and I/Si interfaces [18]. For example, the $M-H_{//}$ curves presented in Sec. II-C reveal that the magnetic properties of the Fe thin films change significantly, depending on the Fe/MgO and Fe/MgAl$_2$O$_4$ interfaces. Based on these considerations, in this section, we study the spin and electron transport properties of a type I structure in a step-by-step manner while varying $t_{MgO}$ and $t_{MAO}$. By comparing the characteristics of different junction structures, crucial parameters for spin injection and a comprehensive analysis method are found accordingly.

## IV-A. Spin injection and $I$-$V$ characteristics in Fe/MgAl$_2$O$_4$/$n^+$-Si tunnel junctions

Here, we originally demonstrate a clear 3TH signal achieved in a type I structure with $t_{MgO} = 0$ and $t_{MAO} \neq 0$, namely, Fe/MgAl$_2$O$_4$/$n^+$-Si structures. Figure 6(a) shows the $I$-$V$ characteristics measured at 10 K for the device with $t_{MAO} = 0.5$ nm and $t_{MgO} = 0$ nm, where the junction area $A = 250$ μm$^2$. Owing to the extremely thin MgAl$_2$O$_4$ layer, the $I$-$V$ curve exhibits a nearly straight line with slight nonlinearity in the $V_B$ range from $-0.5$ V to $-1.5$ V. For the same device, the blue and black curves in Fig. 6(b) show the N-3TH signal measured at 10 K and its fitting curve under $I_B = -100$ mA, where $P_S$ and $\tau_S$ were estimated to be 3.1% and 4.3 ns, respectively. Note that spin injection has never been reported in Si-based tunnel junctions with such a low $RA$ value of 575 Ω·μm$^2$. We also observed clear N-3TH signals with various $I_B$ values, which confirms that the Fe/MgAl$_2$O$_4$(0.5 nm)/$n^+$-Si structure is applicable to the spin injector/detector [see Sec. S2 in S.M[29]]. However, N-3TH signals were not observed as the MgAl$_2$O$_4$ layer became thicker, for example at $t_{MAO} = 0.9$ nm [see Sec. S3 in S.M[29]]. Although the precise physical mechanism remains unclear, this can be attributed to unoptimized interfacial conditions between Fe and MgAl$_2$O$_4$ layers, e.g. the reduced $M_S$ and $M_r$ values of the pink curve observed in Fig. 4. Owing to this unresolved obstacle, we will hereafter focus on 3TH measurements for a type I structure with nonzero $t_{MAO}$ and nonzero $t_{MgO}$ values, since the reduction of $M_S$ and $M_r$ in the Fe thin film was not observed in the type VII structure, and the smooth bilayer barrier is expected to enable efficient spin injection even at low $RA$ values.

## IV-B. Electrical properties of the Fe/MgO/MgAl$_2$O$_4$/$n^+$-Si tunnel junctions

We performed $I$-$V$ measurements at 10 K for type I devices with $A = 250$ μm$^2$ and various combinations of nonzero $t_{MAO}$ and $t_{MgO}$. Figure 7 shows the schematic layout of the 16 combinations of $t_{MAO}$ and $t_{MgO}$ examined in the following sections, where $t_{MAO} = 1.4, 4.1, 5.9,$ and 7.6 Å, and $t_{MgO} = 3.3, 5.5, 7.7,$ and 9.9 Å. As described in Sec. II-A, these different $t_{MAO}$ and $t_{MgO}$ conditions were realized with a shutter moving during a single deposition process, which ensures that the Fe/MgO and MgAl$_2$O$_4$/$n^+$-Si interfaces have the identical characteristics across all the 16 combinations. For clarity, a numbering system from No.1 to 16 is used to designate each combination.

To investigate the electrical properties of the MgO/MgAl$_2$O$_4$ bilayer-based tunnel junctions, several typical $I$-$V$ curves for No.1-No.7, No. 9, No.10, and No.14 combinations are plotted in Fig. 8(a), where the correspondence between the numbers and curves is indicated by the same color. Nonlinearities of these $I$-$V$ curves indicate that the MgO/MgAl$_2$O$_4$ bilayers serve as a tunnel barrier. It is found that the $I$-$V$ characteristics are predominantly determined by $t_{ox}$. For example, at a fixed $I_B = 20$ mA, $V_B = 0.042$ V, 0.127V, and 0.297V were obtained for No.1 ($t_{ox} = 4.7$ Å),

No.2 ($t_{ox}$ = 7.4 Å), and No.3 ($t_{ox}$ = 9.2 Å) combinations, respectively, reflecting an increase in junction resistance with increasing $t_{ox}$. On the other hand, for any two combinations with similar $t_{ox}$ values, a combination with a thicker MgAl$_2$O$_4$ layer results in a slightly higher junction resistance compared with the ones with a thicker MgO layer. For example, No. 3 and No. 9 combinations with similar $t_{ox}$ values of 9.2 Å ($t_{MAO}$ = 5.9 Å/$t_{MgO}$ = 3.3 Å) and 9.1 Å ($t_{MAO}$ = 1.4 Å/$t_{MgO}$ = 7.7 Å) exhibit $V_B$ = 0.297 and 0.204 V at a fixed $I_B$ = 20 mA, respectively. In other words, the No. 3 combination exhibits a higher junction resistance than the No. 9 combination. These features are relevant to the fact that a thick MgAl$_2$O$_4$ layer leads to higher resistance than a thick MgO layer at similar $t_{ox}$ values, and that the MgAl$_2$O$_4$ and MgO layers can play slightly different roles for tunneling electrons, such as exhibiting different effective tunneling masses.

To check whether leakage current is significant in our tunnel junctions or not, we estimated $R_J$ values from the $I$-$V$ curves and analyzed the junction area $A$ dependence of the $R_J$ − $V_B$ curves. For this purpose, No.10 and No. 1 combinations were selected, since they correspond to moderately thick and the thinnest bilayer conditions, respectively. The purple, red, orange, and green curves in Fig. 8(c) show the $R_J$ − $V_B$ curves at 10 K for No.10 combination with $A$ = 25, 250, 2500, and 25000 μm$^2$, respectively. The nearly tenfold scaling of $A$ corresponds to a tenfold change in $R_J(V_B)$, indicating that leakage current is negligibly small for the tunnel junctions with a moderately thick bilayer. On the other hand, the solid brown, black, and gray curves in Fig. 8(d) show the $R_J$ − $V_B$ curves at 10 K for No.1 combination with $A$ = 25, 250, and 25000 μm$^2$, respectively, where tenfold scaling is not well satisfied. This result arises from the fact that the resistance of the $n^+$-Si substrate is not negligible, since the tunnel barrier is extremely thin ($t_{ox}$ = 4.7 Å). For example, the gray curve with $A$ = 25000 μm$^2$ exhibits an almost constant resistance of ~1 Ω that corresponds to the substrate resistance. To eliminate the influence of the $n^+$-Si resistance, we subtracted 1 Ω from the solid brown and black curves, as the dashed brown and black curves shown in Fig. 8(d). After this correction, the tenfold scaling is obtained. Thus, leakage current is negligible small for all the devices, including the junction with the thinnest tunnel barrier.

Finally, we plotted $RA$ values around $V_B$ = 0 V (denoted as $RA_0$) against $t_{ox}$ for all 16 combinations (No. 1 to No.16), as shown by the blue dots in Fig. 8(b). It is found that these data points can be well fitted by a straight line log($RA_0$) ∝ $t_{ox}$ without a significant deviation, from which an effective barrier height $\Phi_B$ ~1.7 eV was estimated [30]. This reveals that the tunneling current is dominated by the direct tunneling mechanism, and that the MgO/MgAl$_2$O$_4$ bilayer can be treated as an effective single barrier layer with a fixed $\Phi_B$ value regardless of the $t_{MAO}$ and $t_{MgO}$ combinations. Noted that $\Phi_B$ ~1.7 eV is significantly higher than that of a MgO single barrier of ~0.3 eV in Fe/MgO/Fe structures [31]. One possible origin is the oxygen defects in the as-deposited MgO layer were compensated by oxygen atoms from the MgAl$_2$O$_4$ layer.

**IV-C. Spin injection in Fe/MgO/MgAl$_2$O$_4$/$n^+$-Si tunnel junctions**

To study how $P_S$ changes with various $t_{MAO}$ and $t_{MgO}$ combinations, 3TH signals were measured for all the 16 combinations, as shown in Fig. 7 with $A$ = 250 μm$^2$. To exclude the bias-dependent influence on $P_S$, different $I_B$ values were applied for different devices so that $V_B$ keeps a nearly constant value of −0.52 V. The selection of $V_B$ = −0.52 V has the following reasons:
(i) For the most of the combinations, the current density $J_B$ = $I_B$/$A$ satisfying $V_B$ = −0.52 V is sufficient to generate an N-3TH signal with a high amplitude and a high signal-to-noise ratio (SNR).
(ii) $V_B$ = −0.52 V is a moderate voltage drop that results in an electric field typically smaller than 7 MV/cm through the MgO/MgAl$_2$O$_4$ bilayer. Thus, tunneling paths generated by a high electric field, such as Fowler-Nordheim (F-N) tunneling, can be excluded in the following analysis [32].
(iii) $P_S$ values estimated under this bias are representative for each combination. It was experimentally confirmed

that $P_S$ is relatively insensitive to small deviations in $V_B$ (less can ±0.1 V) from 0.52V [see Sec. S4 in [29]), which is supported by our theory, i.e., the plateau of $P_{det}$ and moderate change of $P_{inj}$ around −0.5 V [19].

Figure 9(a) shows the typical N-3TH signals measured at 10 K in the spin extraction geometry ($I_B < 0$). The red, orange, green, and purple curves represent the experimental $\Delta V_{3T}$ signals measured for No.10 ($t_{ox}$ =11.8 Å), No.7 ($t_{ox}$ =11.4 Å), No.6 ($t_{ox}$ =9.6 Å), and No.2 ($t_{ox}$ =7.4 Å) combinations, respectively, while the black curves show the corresponding fitting results calculated using Eq. (1). The measurements were performed with $I_B$ = −10, −10, −35, and −65 mA for the red, orange, green, and purple curves, respectively. The estimated values of $P_S$ are 33.5 ± 0.8 %, 27 ± 1.3 %, 13.6 ± 0.3 % and 5.5 ± 0.3 %, and the corresponding $\tau_S$ values are 2.9, 5.8, 3.5, and 2.3 ns, respectively, where the error margins of $\tau_S$ in the fittings are typically smaller than 0.5 ns. It is notable that $P_S$ = 33.5% exceeds the highest value (~25% at 4K) for Fe/MgO/$n^+$-Si junctions [18]. This indicates that $MgAl_2O_4$ provides an effective buffer layer for a smooth MgO thin film, while it reduces the interface defect density compared with the MgO/$n^+$-Si interface, thereby enhancing the $P_S$ value [18]. The estimated $\tau_S$ values in the range from 2 to 6 ns are consistent with the previous reports in $n^+$-Si [11,16,17,21,33,34], which is proof for the spin injection into the $n^+$-Si. All these results find that $t_{ox}$ is the crucial parameter in determining $P_S$: For example, $P_S$ decreases from 27% to 13.6% as $t_{ox}$ is decreased from 11.4 Å to 9.6 Å, as seen by comparing the orange (No. 7) and green (No.6) curves. A similar trend is observed between the green (No.6) and purple (No.2) curves, where $P_S$ decreases from 13.6% to 5.5% as $t_{ox}$ is decreased from 9.6 Å to 7.4 Å. On the other hand, for the devices with similar $t_{ox}$ values, slight variations in $P_S$ are observed depending on some $t_{MAO}$ and $t_{MgO}$ combinations. For instance, the red (No. 10) and orange (No. 7) curves, corresponding to similar $t_{ox}$ values of 11.8 Å ($t_{MAO}$ = 4.1 Å/$t_{MgO}$ = 7.7 Å) and 11.4 Å ($t_{MAO}$ = 5.5 Å/$t_{MgO}$ = 5.9 Å), exhibit different $P_S$ values of 33.5% and 27%, respectively. Thus, our finding is that $MgAl_2O_4$ and MgO contribute differently to the $P_S$ values. Nevertheless, as will be shown later, such differences are significantly smaller than the overall dependence of $P_S$ on $t_{ox}$. Therefore, we will primarily focus on the $t_{ox}$ dependence of $P_S$ in the following analysis.

## V. $P_S − t_{ox}$ and $P_S − RA_{PS}$ relationships in Fe/MgO/MgAl$_2$O$_4$/$n^+$-Si tunnel junctions

The analysis of the $I−V$ characteristics and 3TH signals for the type I structure with various $t_{MAO}$ and $t_{MgO}$ combinations in Sec. IV-C reveals that both electron and spin transport are predominantly determined by $t_{ox}$. In other words, $t_{ox}$ is the crucial parameter for the analysis of $P_S$. In this section, we systematically study the relationship between $P_S$ and $t_{ox}$ across a wide range of $t_{ox}$, based on the combinations examined in Sec. IV-C. It is found that the features of the $P_S − t_{ox}$ plot are distinctive and cannot be explained by the conventional linear conductivity-matching model for Fe/I/Si structures [26−28]. Furthermore, we show a more general understanding of the $P_S − t_{ox}$ plot when $t_{ox}$ is converted into $RA$. This approach paves the way for qualitative interpretations of spin transport, as described in the following sections.

### V-A. Experimental $P_S − t_{ox}$ plot

The blue dots in Fig. 9(b) show the estimated $P_S$ values plotted against $t_{ox}$ for various $t_{MAO}/t_{MgO}$ combinations. Among the 16 combinations, No. 12, 15, and 16 were excluded in the plot due to their extremely low SNR. The blue rectangle is the data point estimated from the results for the Fe/MgAl$_2$O$_4$(0.5 nm)/$n^+$-Si junction in Figs. 6. As $t_{ox}$ is increased from 2 to 11 Å, $P_S$ steeply increases in a monotonic manner. However, with a further increase of $t_{ox}$ beyond 11 Å, $P_S$ fluctuates around 30% without any significant further rise. The features directly show that $P_S$ and $t_{ox}$ are physically corelated, and strongly suggest that the spin transport process can be interpreted by two distinct physical mechanisms that dominate in the thin and thick $t_{ox}$ ranges, respectively. For each data point in Figure 9(b), the junction resistance-area product $RA_{PS}$ (= $A \times V_B / I_B$), at which $P_S$ was estimated, was plotted as a function of $t_{ox}$, as shown in Fig. 9(c). Similar to the $RA_0$–$t_{ox}$ plot shown in Fig. 8(a), log($RA_{PS}$) increases with $t_{ox}$ in a nearly linear manner as indicated by the red dashed line. This confirms that the direct tunneling mechanism is still dominant under the bias conditions. To gain a clear physical insight into the spin transport process as $t_{ox}$ is changed, we transformed the $P_S$ − $t_{ox}$ plot into the $P_S$ − $RA_{PS}$ plot using the $RA_{PS}$ − $t_{ox}$ plot, as the blue dots shown in Fig. 9(d). This transformation can significantly facilitate our analysis for the following three reasons: (i) The realistic junction properties are more accurately represented by the $P_S$ − $RA_{PS}$ plot than the $P_S$ − $t_{ox}$ plot. This is because the error margins of the estimated $t_{ox}$ values (approximately ±1 Å) can largely affect the features of the $P_S$ − $t_{ox}$ plot. In contrast, the electrically-estimated $RA_{PS}$ values have much higher accuracy and include realistic junction properties like the roughness of the barrier. (ii) To provide quantitative explanations, it is more straightforward to establish a physical relationship between $P_S$ and $RA_{PS}$ than that between $P_S$ and $t_{ox}$. This is because both $P_S$ and $RA_{PS}$ are electrically defined by tunneling current [19], whereas $t_{ox}$ represents a material property. (iii) The $P_S$–$RA_{PS}$ plot offers a unified approach to understanding and studying the barrier thickness dependence of the spin polarization. It will be shown later that our $P_S$ − $RA_{PS}$ plot demonstrates a certain level of generality when compared with Ref. [16,18,23].

**V-B. Experimental $P_S$ − $RA_{PS}$ plot**

Hereafter we focus on the analysis of the $P_S$–$RA_{PS}$ plot shown in Fig. 9(d). As $RA_{PS}$ is increased from 600 to 13000 Ω·μm$^2$, $P_S$ steeply increases from 0 to 30 (±5) % in a nearly linear manner, which corresponds to the monotonic increasing trend in the $P_S$ − $t_{ox}$ plot when $t_{ox}$ is increased from 2 to 11 Å. As $RA_{PS}$ is further increased from 13000 to 50000 Ω·μm$^2$, $P_S$ does not steeply increase, but is nearly constant at 30% with a deviation of ±5 %, which corresponds to the fluctuation region in the $P_S$ − $t_{ox}$ plot when $t_{ox}$ > 11 Å. It is worth noting that $P_S$–$RA_{PS}$ characteristics across such an extensive $RA_{PS}$ range have never been observed with the same kind of junction structures having the identical interface quality. Furthermore, our $P_S$ − $RA_{PS}$ plot demonstrates a degree of generality applicable to the previous studies on the Fe/I/$n^+$-Si structures. The colored squares in Fig. 9(d) show the experimental ($P_S$, $RA_{PS}$) data points extracted from literature [16,18,23] and Figs. 6(a) and (b), where the corresponding junction structures and measurement temperatures are denoted. As $RA_{PS}$ increases, these data points closely follow the rising trend of our $P_S$ − $RA_{PS}$ plot, regardless of the different junction structures. This result indicates a similar tunneling mechanisms that govern the spin transport process through Fe/I/$n^+$-Si junctions.

It should be noted that the $RA_{PS}$ dependence of $P_S$ cannot be explained by the well-known V-F model proposed for conductivity matching problem [28], where $P_{inj}$ increases steeply and then saturates as the tunnel barrier resistance $r_b$ becomes much larger than the Si spin resistance $r_{S,Si}$. In this study, the $r_{S,Si}$ for $n^+$-Si is significantly smaller than the $r_b$ values for all the combinations. For example, the No.1 combination that has the smallest $r_b$ value (~600 Ω·μm$^2$) is much larger than the $r_{S,Si}$ value of $n^+$-Si (~30 Ω·μm$^2$) [16]. Our numerical calculations have confirmed that the conductivity mismatch formula [28] cannot account for the experimental $P_S$ − $RA_{PS}$ trend [see Fig. S5 in Section S5 in S.M. [29]]. Therefore, other physical mechanisms are needed to explain our $P_S$ − $RA_{PS}$ plot.

**VI. Proposal of a phenomenological "two-path model" and its fitting to the experimental $P_S$ − $RA_{PS}$ plot**

The features of our $P_S$ − $RA_{PS}$ plot indicate that the spin transport is governed by different mechanisms in the high and low $RA_{PS}$ ranges, respectively, while they do not originate from the conventional linear conductivity-matching model for Fe/I/Si structures [28]. Besides, our previous work revealed that the conventional linear model is inadequate for describing the spin transport in Fe/I/Si structures, particularly under the present conditions where $V_B$ is significantly higher than 0V [19]. Thus, in this section, we first compare the experimental $P_S - RA_{PS}$ plot with a theoretical $P_S - RA_{PS}$ curve calculated from our *non-linear* spin transport model. This comparison reveals that the *non-linear* model can partially account for the $P_S - RA_{PS}$ characteristics, but additional mechanisms are needed for a complete explanation. To capture the features of these mechanisms, we propose a phenomenological "two-path model" based on the previous *non-linear* model to perform quantitative fittings to the experimental results. The extracted fitting parameters provide insights into the different spin transport mechanisms, which are dominant in high and low $RA_{PS}$ ranges.

**VI-A. Comparison of the $P_S - RA_{PS}$ plot with our previous theoretical model**

In the high $RA_{PS}$ range (13000 ~ 50000 Ω·μm²), the fluctuation features of $P_S$ can be analyzed by our previous theoretical model based on the direct tunneling mechanism and s-like band structure of Fe and $n^+$-Si [19]:

$$J_\pm(V, \Delta\mu(V), t_{ox}) = A \int_{-\infty}^{+\infty} D_{Fe}^\pm(E,V) \, D_{Si}(E,V) \, T^\pm(E, V, \Delta\mu(V), t_{ox}, \Phi_{eff}, m_t) \, dE, \qquad (2)$$

where $J_\pm$ is the current density for up-spin / down-spin (+/−) electrons, $V$ is the applied voltage bias across the junction, $\Delta\mu(V)$ is the spin accumulation at $V$ on the $n^+$-Si side, $t_{ox}$ is the barrier thickness, $E$ is the electron energy, $D_{Si}$ is the density of states (DOS) in $n^+$-Si, $D_{Fe}^\pm$ are the DOSs for up-spin / down-spin (+/−) electrons in Fe, $\Phi_{eff}$ is the effective barrier height, and $T^\pm$ are the effective up-spin / down-spin (+/−) tunneling probabilities for electrons with a tunneling effective mass $m_t$. Note that $T^\pm$ is related to all the parameters $E$, $V$, $\Phi_{eff}$, $t_{ox}$, and $m_t$. The constant $A = 7 \times 10^{11}$ A/m²/(eV)² was estimated by fitting the experimental $J$–$V$ curve [19]. By setting various $t_{ox}$ values in Eq. (2), we numerically calculated $P_S$ under various $RA_{PS} = V/(J_+ + J_-)$ values at $V = -0.52$ V. The result is shown by the green curve in Fig. 4(d), plotted with a right-side horizontal axis. Here, we focus on the difference of the $RA_{PS}$ dependences of $P_S$ between the experimental data (blue dots) and calculation (green curve), since comparing the absolute values requires realistic material parameters as input to Eq. (2), which are unknown for the amorphous barrier and polycrystalline Fe at present. The green curve shows a little variation as $RA_{PS}$ is increased from 600 to $1\times10^6$ Ω·μm². This agrees with the fluctuation of the blue dots ($P_S = 25 - 35$ %) in the high $RA_{PS}$ range within their error bars, but it does not agree with the steep increase of the blue dots observed in the low $RA_{PS}$ range (600 ~ 13000 Ω·μm²). Hence, additional spin transport mechanisms, distinct from those described by Eq. (2), probably contribute to the reduced $P_S$ values in the low $RA_{PS}$ range. Although the microscopic origins remain unclear, we will construct a phenomenological model to characterize the features of the spin transport in the low $RA_{PS}$ range. For clarity, we hereafter refer to the tunneling path described by Eq. (2) as Path-A with high spin polarization, and an additional tunneling path distinct from Eq. (2) as Path-B with low spin polarization.

**VI-B. Construction of a phenomenological "two-path" model**

Here, we first qualitatively describe the features of Path-A and Path-B using schematic band diagrams. Figures 10 (a) and (b) show the schematic band diagrams of the spin transport process at a constant $V_B$ (= 0.52 V) bias in the low and high $RA_{PS}$ ranges, respectively, where $\bar{\mu}^{Fe}$ and $\bar{\mu}^{Si}$ are the Fermi levels of Fe and $n^+$-Si, respectively, red and blue colors in Fe represent the filled states of up-spin and down-spin electrons, respectively, $J_A$ and $J_B$ are the electron current densities of Path-A and Path-B, and $P_{S,A}$ and $P_{S,B}$ are the electron spin polarizations of Path-A and Path-B, respectively, with $P_{S,A}$ being larger than $P_{S,B}$. The widths of the pink and green arrows express the magnitudes of $J_A$ and $J_B$, respectively. As explained in Sec. V-A, it is reasonable to assume that both Path-A and Path-B follow the direct tunneling mechanism with negligible leakage current through the barrier. Nevertheless, based on the features observed in the $P_S - RA_{PS}$ plot, two paths are inferred to contribute differently to $P_S$ in the low and high $RA_{PS}$ ranges, respectively. Their key characteristics are described below: In the low $RA_{PS}$ range shown in Fig. 10(a), the weakly spin-polarized Path-B dominates the total tunneling current, namely, $J_A < J_B$, which leads to a low $P_S$ value close to $P_{S,B}$. As $RA_{PS}$, namely, $t_{ox}$, is increased, $J_B$ decreased faster than $J_A$. As a result, in the high $RA_{PS}$ range shown in Fig. 10(b), the highly spin-polarized Path-A becomes dominant in the total tunneling current, namely, $J_A > J_B$, which leads to a high $P_S$ value close to $P_{S,A}$. As $RA_{PS}$ is further increased, the $J_B/J_A$ ratio approaches to zero, resulting in the asymptotic convergence of $P_S$ to $P_{S,A}$, which corresponds to the nearly unchanged $P_S$ values in the high $RA_{PS}$ range, as shown by the green curve in Figs. 9(d).

Next, we propose phenomenological formulas for Path-A and Path-B to provide an explanation of the experimental $P_S - RA_{PS}$ and $RA_{PS} - t_{ox}$ plots. Since both paths follow the direct tunneling mechanism, the form of Eq. (2) for Path-A is also applicable to Path-B. However, it is challenging to directly apply Eq. (2) to numerical calculations of the $P_S - RA_{PS}$ and $RA_{PS} - t_{ox}$ curves, because the exact band structures and material parameters are inaccessible for both paths due to the amorphous tunnel barrier and polycrystalline Fe. To capture the main physical features and enable practical calculations for Path-A and Path-B, we simplify Eq. (2) under a constant bias $V_B$ into a more tractable form, which is then used to construct formulas for Path-A and Path-B. The simplification procedures are described as follows:

$$J_{A,\pm}(V_B, \Delta\mu(V_B), t_{ox}) = A \int_{-\infty}^{+\infty} D_{Fe}^{\pm}(E, V_B)\, D_{Si}(E, V_B)\, T^{\pm}(E, V_B, \Delta\mu(V_B), t_{ox}, \Phi_{\text{eff}}(V_B), m_t)\, dE$$

$$= A \times Const_1 \int_{-\infty}^{+\infty} DOS_{Fe}^{\pm}(E)\, DOS_{Si}(E)\, T_A^{\pm}(E, t_{ox})\, dE$$

$$= A \times Const_1 \int_{-\infty}^{+\infty} DOS_A^{\pm}(E)\, T_A^{\pm}(E, t_{ox})\, dE$$

$$= A \times Const_1 \times Const_2 \times DOS_A^{\pm} \times T_A^{\pm}(t_{ox})$$

$$J_{A,\pm}(t_{ox}) = C_A \frac{1 \pm P_{S,A}}{2} DOS_A \exp\left(-\frac{t_{ox}}{\lambda_A}\right). \tag{3}$$

In the second line, we extract all constant parameters ($V_B$, $\Delta\mu(V_B)$, $\Phi_{\text{eff}}(V_B)$, and $m_t$) from the integral as a constant term $Const_1$, leaving the functional dependencies on $E$ as $DOS_{Fe}^{\pm}(E)$, $DOS_{Si}(E)$, and $T_A^{\pm}(E, t_{ox})$. In the third line, $DOS_{Fe}^{\pm}(E)$ and $DOS_{Si}(E)$ are combined into a single term $DOS_A^{\pm}(E)$. In the fourth line, we assume that the integral can be approximated by $Const_2 \times DOS_A^{\pm} \times T_A^{\pm}(t_{ox})$, where $DOS_A^{\pm}$ is a two-valued constant depending on the spin channel (+ or −), and $T_A^{\pm}(t_{ox})$ is a function of $t_{ox}$. We also assume that $T_A^{\pm}(t_{ox})$ can be expressed in an exponential form, $\exp\left(-\frac{t_{ox}}{\lambda_A}\right)$, by factoring out constants. Finally, Eq. (3) is obtained by collecting all constant terms into $C_A$ and representing spin-related terms as $\frac{1 \pm P_{S,A}}{2}$, where $J_{A,\pm}$ is the current density of Path-A for up-spin / down-spin (+/−) electrons, $P_{S,A}$ is the spin polarization of Path-A, $DOS_A$ is a dimensionless parameter proportional to the product of available initial electron states and final empty states that contribute to $J_A$, $\lambda_A$ is the tunneling decay length of Path-A, and $C_A$ is a constant with the unit of A/m$^2$. It is worth noting that the above simplification is *not* mathematically rigorous. To validate its effectiveness, we performed numerically calculations of the $J_A - t_{ox}$ curves under $V_B = 0.52$ V using either Eq. (2) or Eq. (3), as detailed in [see S6 in S.M.[29]]. Our results confirm that Eq. (3) is a good approximation to Eq. (2) for discussing the $J_A - t_{ox}$ and $P_S - RA_{PS}$ characteristics at a fixed $V_B$. By utilizing this more tractable Eq. (3), we can formally construct the formula for Path-B:

$$J_{B,\pm}(t_{ox}) = C_B \frac{1 \pm P_{S,B}}{2} DOS_B \exp(-t_{ox}/\lambda_B), \tag{4}$$

where $J_{B,\pm}$ is the current density of Path-B for up-spin / down-spin (+/−) electrons, $P_{S,B}$ is the spin polarization of Path-B, $DOS_B$ is a dimensionless parameter proportional to the product of available initial electron states and final empty states that contribute to $J_B$, and $\lambda_B$ is the tunneling decay length of Path-B, and $C_B$ is a constant with the unit of A/m².

### VI-C. Fitting the experimental $P_S - RA_{PS}$ and $RA_{PS} - t_{ox}$ plots by a phenomenological "two-path" model

By using Eqs. (3) and (4), the $P_S - RA_{PS}$ and $RA_{PS} - t_{ox}$ characteristics were simultaneously calculated and fitted to the experimental plots with a common set of fitting parameters. First, the total electron current $J_{total}$ and spin current $J_{S,total}$ contributed by Path-A and Path-B are defined as follows:

$$J_{total}(t_{ox}) = \left(J_{A,+}(t_{ox}) + J_{A,-}(t_{ox})\right) + \left(J_{B,+}(t_{ox}) + J_{B,-}(t_{ox})\right), \tag{5}$$

$$J_{S,total}(t_{ox}) = \left(J_{A,+}(t_{ox}) + J_{A,-}(t_{ox})\right) - \left(J_{B,+}(t_{ox}) + J_{B,-}(t_{ox})\right). \tag{6}$$

Then, $P_S$ and $RA_{PS}$ are calculated using:

$$P_S \sim P_{inj} = \frac{J_{S,total}(t_{ox})}{J_{total}(t_{ox})}, \tag{7}$$

$$RA_{PS} = \frac{|V_B|}{J_{total}(t_{ox})} + RA_{para}, \tag{8}$$

where $|V_B| = 0.52$ V is a constant, and $RA_{para} = 250$ Ω·µm² accounts for the parasitic resistance-area product from the $n^+$-Si substrate, which has a resistance of approximately 1 Ω (see the gray curve in Fig. 8(d)). Equation (7) serves as a good approximation to $P_S = \sqrt{P_{inj}P_{det}}$, since $P_{inj}$ and $P_{det}$ exhibit a nearly linear relationship with a proportionality factor close to 1 [19]. It is noted that this fitting method only allows estimation of the products $C_B \times DOS_B$ and $C_A \times DOS_A$; that is, the absolute values of $C_{A(B)}$ and $DOS_{A(B)}$ cannot be independently determined. Since our interest lies in the ratio $DOS_B/DOS_A$, we assume $C_B = C_A$ and normalize $DOS_A = 1$ in the following calculations. The blue dashed lines in Figs. 11(a) and (b) show the best-fitting curves to the experimental $P_S - RA_{PS}$ and $RA_{PS} - t_{ox}$ data extracted from Figs. 9(d) and (c), respectively. These fittings were obtained by simultaneously minimizing the total root-mean-square deviation (RMSD) values across both plots [see Sec. S7 in S.M.]. The good agreement between the experimental data and calculations confirms the validity of introducing Path-B. The obtained best fit parameters are consistent with the qualitative features of Path-A and Path-B illustrated in Figs. 10(a) and (b): Path-A is characterized by a high spin polarization $P_{S,A} = 32\%$ and a long decay length $\lambda_A = 0.44$ nm, while Path-B exhibits a low spin polarization $P_{S,B} = 0\%$ and a short decay length $\lambda_B = 0.1$ nm. The proportionality constant was determined to be $C_B = C_A = 3.23 \times 10^{-4}$ A/µm². In addition, $DOS_B$ was found to be larger than $DOS_A$ by a factor of 1500, which suggests that Path-B provides more available states contributing to the tunneling current. To evaluate the robustness of the fitting, we define a 20% increase in the minimum RMSD value as the criterion for reasonable parameter variation, under the fixed $P_{S,A} = 32\%$ and $P_{S,B} = 0\%$. Within this range, the fitting yields the following parameter intervals: $\lambda_A = 0.33 \sim 0.47$ nm, $\lambda_B = 0.084 \sim 0.12$ nm, and $DOS_B/DOS_A = 200 \sim 10000$ [see Sec. S7 in S.M.[29]].

## VII. Discussion

The fitting results presented in Sec. VI revealed that the spin transport in the Fe/MgO/MgAl₂O₄/$n^+$-Si junctions can be well explained by two phenomenological tunneling paths with distinct features: A highly spin-polarized Path-A with a long tunneling decay length $\lambda_A$ and smaller effective density of states $DOS_A$, and a weakly spin-polarized

Path-B with a short tunnel decay length $\lambda_B$ and larger effective density of states $DOS_B$. These characteristics determine that Path-A and Path-B dominate the tunneling current and $P_S$ value in the high and low $RA_{PS}$ ranges, respectively, which lead to the steep increase in $P_S$ as $RA_{PS}$ (or $t_{ox}$) is increased in our experiments. Since Path-A is simplified from our previous model, it corresponds to direct tunneling between the s-like bands of Fe and Si. On the other hand, Path-B is phenomenologically constructed based on experimental results, and its physical origins remain unclear. To gain a deeper insight into the physical origin of Path-B, this section discusses potential mechanisms using supplementary experimental analyses and qualitative comparison of the obtained fitting parameters with theoretical predictions for coherent tunneling.

**VII-A. Analysis of magnetically dead layer at the Fe/MgO interface**

In this section, we examine whether the suppression of $P_S$ in the thin $t_{ox}$ range, i.e., the contribution from Path-B, originates from the formation of magnetically dead layers (MDLs) at the Fe/MgO interfaces. Here, an MDL refers to as an ultrathin (one atomic layer or less) paramagnetic layer or paramagnetic interface states, which can significantly reduce $P_S$ depending on their specific magnetic properties [15,16]. In our previous work, it was demonstrated that such reduction can be resolved through the optimization of an insertion Mg layer inserted between Fe and MgO [16]. Here, we study whether the influence of MDLs on $P_S$ becomes pronounced as $t_{ox}$ is decreased under a fixed Mg thickness.

The influence of MDLs on $P_S$ can be qualitatively evaluated by the analysis of the ratio between the magnitudes of B-3TH and I-3TH signals, defined as follows [15]:

$$\frac{M[\Delta V_{3T}^I]}{M[\Delta V_{3T}^B]} = M\left[\frac{\Delta V_{3T}^I}{\Delta V_{3T}^B}\right] \propto \frac{\sum_i (H_{\text{Fe-perp},i}^{\text{MDL}})^2}{(H_{\text{Fe-paral}}^{\text{MDL}})^2}, \tag{9}$$

where $M[\Delta V_{3T}^I]$ and $M[\Delta V_{3T}^B]$ are the magnitudes of B-3TH and I-3TH signals fitted by Lorentzian function, respectively, and $H_{\text{Fe-paral}}^{\text{MDL}}$ and $H_{\text{Fe-perp},i}^{\text{MDL}}$ denote the MDL magnetic fields parallel and perpendicular to the Fe magnetization direction, respectively. In the 3TH measurements, the transport of spin-polarized electrons is suppressed by $H_{\text{Fe-perp},i}^{\text{MDL}}$, and enhanced by $H_{\text{Fe-paral}}^{\text{MDL}}$, respectively. Thus, a small ratio of $M\left[\frac{\Delta V_{3T}^I}{\Delta V_{3T}^B}\right]$ leads to a high $P_S$ value. Figures 12(a–c) show the measured $\Delta V_{3T}^I$ (top) and $\Delta V_{3T}^B$ (bottom) signals at 10 K, obtained under $H_{//}$ and $H_\perp$ sweeps, respectively, with a fixed $V_B = 0.52$ V. The red, green, and brown curves represent the experimental $\Delta V_{3T}$ signals measured for No.10 ($t_{ox}$ =11.8 Å), No.6 ($t_{ox}$ =9.6 Å), and No.5 ($t_{ox}$ =6.9 Å) combinations, respectively, while the black curves show the corresponding Lorentzian fitting results. Their corresponding $M[\Delta V_{3T}^I/\Delta V_{3T}^B]$ values were estimated to be 0.16, 0.074, and 0.084, respectively. It is found that the $M[\Delta V_{3T}^I/\Delta V_{3T}^B]$ values for No. 6 and No. 5 combinations are smaller than that of No. 10, indicating that the influence of MDLs on $P_S$ becomes less pronounced in the thicker $t_{ox}$ range. Therefore, MDLs are not responsible for the reduction of $P_S$ in the thin $t_{ox}$ range, namely, the physical origin of Path-B.

**VII-B. Analysis of temperature dependence of junction $RA$ values**

In this section, we investigate whether Path-B originates from hopping conductance via defects within the MgO/MgAl$_2$O$_4$ bilayer or not. Here, defects refer to localized states with randomly oriented spins, and hopping conduction (HC) through such states is generally regarded as a mechanism that relaxes spin-polarized currents [35]. As analyzed in Sec. IV-B, the nearly exponential dependence of $RA_0$ on $t_{ox}$ indicates that direct tunneling (DT) is the dominant transport mechanism in our junctions. However, since HC can also exhibit an exponential dependence on $t_{ox}$, the $RA_0 - t_{ox}$ relationship alone does not rule out the possibility that HC significantly contributes to the tunneling current, thereby suppressing $P_S$. Thus, it necessary to examine the influence of HC on $RA_0$ for various $t_{ox}$ values and assess whether HC becomes prominent at smaller $t_{ox}$. HC typically includes two mechanisms; elastic hopping conduction (EHC), which is temperature-independent, and inelastic hopping conduction (IHC), which depends on temperature. In the following, we first analyze IHC by studying the temperature dependence of the junction conductance for various $t_{ox}$. Then, we qualitatively assess the relevance of EHC to Path-B by comparing theoretical models of DT and EHC.

Figure 13 shows the conductance per area, $\sigma_0 = 1/RA_0$, plotted as a function of temperature $T$. The red, purple, and brown open dots represent the experimental data for combinations No.10 ($t_{ox}$ =11.8 Å), No.2 ($t_{ox}$ =7.4 Å), and No.5 ($t_{ox}$ =6.9 Å), respectively. As $T$ increases, a moderate rise in $\sigma_0$ is observed for all three combinations, indicating that the EHC is not negligible at elevated temperatures. To further clarify the underlying physics of the $T$-dependence of $\sigma_0$, it is necessary to determine whether EHC becomes negligible at 10K. The experimental $\sigma_0-T$ data were fitted using the following formula [35]:

$$\sigma_0(T) = \sigma_{DT} + \sigma_{EHC-1} + \sum_{n \geq 2} \sigma_{IHC,n} = \sigma_{DT-EHC} + \sum_{n \geq 2} \alpha_n T^{\beta_n}, \tag{10}$$

where $\sigma_{DT}$ is the $T$-independent conductance per area due to DT, $\sigma_{EHC-1}$ is the EHC per area via a single localized state, and $\sigma_{DT-EHC}$ is defined as the sum of $\sigma_{DT}$ and $\sigma_{EHC-1}$. The term $\sigma_{IHC,n}$ represents the IHC per area via chains of $n \geq 2$ localized states, modeled as $\sigma_{IHC,n} = \alpha_n T^{\beta_n}$, with $\alpha_n$ and $\beta_n$ as fitting parameters. IHC involving a single localized state ($n = 1$) is excluded, as it is regarded as a minor correction to $\sigma_{EHC-1}$ and negligibly small [35−37]. Similarly, EHC contributions from chains with $n \geq 2$ are omitted in Eq. (10), because their impact is extremely limited compared with that of IHC for the same $n \geq 2$ [35]. Theoretical studies have shown that $\sigma_{IHC,n}$ exhibits a characteristic $T$ dependence depending on $n$, given by [35,37]:

$$\sigma_{IHC,n} \propto T^{\frac{n^2+n-2}{n+1}}. \tag{11}$$

For example, the exponents $\beta_n$ = 4/3 and 5/2 correspond to $\sigma_{IHC,n}$ with $n$ = 2 and 3, respectively. These theoretical values serve as benchmarks for fitting to the experimental data. In Fig. 13, the red, purple, and brown dashed curves show the fitting results obtained using Eq. (10) for No.10 ($t_{ox}$ =11.8 Å), No.2 ($t_{ox}$ =7.4 Å), and No.5 ($t_{ox}$ =6.9 Å) combinations, respectively, where the RMSD between the fitting curves and experimental data was minimized to obtain the best-fitting curves. The corresponding fitting parameters are listed in Table 1. For No.2 and No.5 combinations with thin $t_{ox}$, the best fittings were obtained by only the $\sigma_{IHC-2}$ term. The estimated $\beta_2$ values were found to be 1.29 and 1.33, respectively, which are in close agreement with the theoretical prediction $\beta_2$ = 4/3. This consistency indicates that IHC via chains of two localized states is the dominant mechanism contributing to the increase in $\sigma_0$ at high temperatures. Unlike this result, for No.10 combination with thick $t_{ox}$, a fitting using only $\sigma_{IHC-2}$ does not yield physically meaningful parameters. Thus, $\sigma_{IHC-3}$ is included in Eq.(10) while fixing $\beta_2$ = 1.33. In this manner, the best fitting produced a $\beta_3$ value of 2.62, consistent the theoretical prediction $\beta_3$ = 5/2, indicating that IHC via chains of three localized states dominates the high-temperature conductance behavior.

To assess the significance of IHC at 10 K across different $t_{ox}$, the values of $\sigma_0$ at $T$ = 0 were extrapolated from the fitted curves. The ratios $\sigma_0(T=0)/\sigma_0(T=10\text{ K})$ for each combination are listed in the last column of Table 1. It is found that the $T$-independent component $\sigma_{DT-EHC}$ accounts for more than 97% of $\sigma_0$ for all three combinations at 10 K, indicating that IHC is negligible small in our junctions across a wide range of $t_{ox}$. Therefore, IHCs do not account for the physical origin of Path-B.

|  | $\sigma_{\text{DT-EHC}}$ (S/m²) | $\alpha_2$ (S/m²) | $\beta_2$ | $\alpha_3$ (S/m²) | $\beta_3$ | $\sigma_0(T=0)/\sigma_0(T=10\text{ K})$ |
|---|---|---|---|---|---|---|
| No.2 ($t_{\text{ox}}$ =7.4 Å) | 4.2×10⁸ | 4.9×10⁶ | 1.29 |  |  | 97.7% |
| No.5 ($t_{\text{ox}}$ =6.9 Å) | 1.9×10⁸ | 1.4×10⁶ | 1.40 |  |  | 98.2% |
| No.10 ($t_{\text{ox}}$ =11.8 Å) | 1.7×10⁶ | 3.7×10² | 1.33 (fixed) | 2.34 | 2.62 | 99.5% |

Table 1 Estimated fitting parameters $\sigma_{\text{DT-EHC}}$, $\alpha_2$, $\beta_2$, $\alpha_3$, and $\beta_3$ using Eq. (10) for No.2, No.5, and No.10 combinations. The last column lists the calculated ratios $\sigma_0(T=0)/\sigma_0(T=10\text{ K})$ for each combination.

The analysis of the $T$ dependence of $\sigma_0$ presented above does not allow us to distinguish the relative contributions of $\sigma_{\text{EHC-1}}$ and $\sigma_{\text{DT}}$, since both are temperature-independent and contribute to $\sigma_0$ even at 0 K. Nevertheless, the distinct theoretical $t_{\text{ox}}$ dependences of $\sigma_{\text{EHC-1}}$ and $\sigma_{\text{DT}}$ offer an alternative approach to evaluate whether $\sigma_{\text{EHC-1}}$ also can account for the physical origin of Path-B. The $t_{\text{ox}}$ dependences of $\sigma_{\text{DT}}$ and $\sigma_{\text{EHC-1}}$ are theoretically given by [35]:

$$\sigma_{DT} \propto \exp\left(2\frac{t_{\text{ox}}}{\lambda_0}\right), \tag{12}$$

$$\sigma_{EHC-1} \propto \exp\left(2\frac{(t_{\text{ox}}/2)}{\lambda_0}\right) = \exp\left(\frac{t_{\text{ox}}}{\lambda_0}\right), \tag{13}$$

where $\lambda_0$ is the localization length of the defect states within the barrier. The factor 2 in Eq. (12) arises from the quantum mechanical treatments of tunneling, such as the WKB approximation. In contrast, the effective barrier width in Eq. (13) is $t_{\text{ox}}/2$, which reflects the fact that hopping conduction is most probable when the localized state is positioned at the center of the barrier [35]. As a result, it is theoretically expected that the effective tunneling decay length for $\sigma_{\text{DT}}$ is $\lambda_0/2$, which is half that of $\sigma_{\text{EHC-1}}$, e.g., $\lambda_0$. In other words, if Path-B originates from the spin relaxation associated with $\sigma_{\text{EHC-1}}$, the extracted decay length $\lambda_B$ is nearly twice as large as that associated with $\sigma_{\text{DT}}$. However, this expectation is qualitatively inconsistent with our fitting results: Path-A that is attributed to the direct tunneling between s-like bands exhibits a decay length $\lambda_A = 0.44$ nm, which is approximately four times larger than $\lambda_B = 0.1$ nm. Based on this contradiction, we therefore conclude that EHCs do not account for the physical origin of Path-B.

**VII-C. Qualitative comparison with coherent tunneling**

The discussions in Secs. VII-A and VII-B rule out magnetically dead layers (MDLs) and hopping conduction (HC) as the physical origins of Path-B. In this section, we examine whether Path-B could originate from the intrinsic band structure of the Fe/MgO/MgAl₂O₄/$n^+$-Si junctions. The analysis is conducted by qualitatively comparing our fitting results with theoretical predictions for coherent tunneling (CT) in epitaxial Fe(001)/Insulator(001)/Fe(001) structures. While the detailed physics remains to be clarified, this comparison provides a possible explanation for the emergence of Path-A and Path-B.

The observed $P_S - t_{\text{ox}}$ characteristics in our experiment resemble the well-known features of tunneling magnetoresistance (TMR) ratio: as an insulator thickness $t_I$ is increased, a sharp increase in TMR ratio followed by saturation has been experimentally observed and theoretically predicted [31,38−41]. These features are attributed to the wavefunction symmetry-dependent tunneling, usually referred to as CT, where the fully spin-polarized $\Delta_1$ band in Fe ($P_{S, \Delta_1} \sim 100\%$) exhibits much longer decay length $\lambda_{\Delta_1}$ than other weakly spin-polarized bands, such as $\Delta_2$, $\Delta_5$ bands. As $t_I$ is increased, the tunneling current becomes increasingly dominated by the $\Delta_1$ band, leading to a pronounced increase followed by saturation in both spin polarization and TMR ratio.

In our junctions, the presence of polycrystalline Fe and amorphous MgO/MgAl₂O₄ bilayer is expected to hinder CT due to structural disorder. Nevertheless, a comparison between our fitting results for Path-A and Path-B and the theoretical predictions for CT through $\Delta_1$ and $\Delta_5$ bands reveals similarity. Table 2 summarizes the theoretical $\lambda_{\Delta 1}$ and $\lambda_{\Delta 5}$ at $k_\parallel = 0$ in the 2D Brillouin zone of the Fe(001) plane, calculated for MgO and MgAl$_2$O$_4$ barriers using first-principles methods [40,42], along with our experimentally extracted fitting parameters $\lambda_A$ and $\lambda_B$ in Fig. 11(a). The corresponding ratios $\lambda_A/\lambda_B$ and $\lambda_{\Delta 1}/\lambda_{\Delta 5}$ are also listed in the last row. It is found that the decay length for Path-A, $\lambda_A = 0.44$ nm, is comparable to the theoretical $\lambda_{\Delta 1}$ values (0.16, 0.24, and 0.30 nm), while $\lambda_B = 0.1$ nm agrees well with the theoretical $\lambda_{\Delta 5}$ values (0.067, 0.063, and 0.13 nm). Moreover, $\lambda_A/\lambda_B = 4.4$ closely matches $\lambda_{\Delta 1}/\lambda_{\Delta 5} = 2.4$, 3.8, and 2.3. The slightly larger $\lambda_A/\lambda_B$ than $\lambda_{\Delta 1}/\lambda_{\Delta 5}$ is reasonable, because $\lambda_B$ may include contributions from other weakly spin-polarized bands like $\Delta_2$, which has a shorter decay length $\lambda_{\Delta 2}$ than $\lambda_{\Delta 5}$ [39,40,42], thus reducing the effective $\lambda_B$. These comparisons suggest that the highly spin-polarized Path-A ($P_{S,A} = 32\%$) and weakly spin-polarized Path-B ($P_{S,B} \approx 0\%$) are associated with tunneling through the $\Delta_1$ band and $\Delta_5$ (or $\Delta_2$) bands, respectively. The structural disorder in the Fe and MgO/MgAl$_2$O$_4$ may explain the reduced spin polarization of Path-A compared with the ideal $P_{S,\Delta 1} = 100\%$. In addition, at the Fe(001)/MgO(001) interface, first-principles calculations have shown a significant enhancement of the minority-spin DOS near the Fe Fermi level, which is attributed to the surface states formed by the Fe $d$-band electrons [44−46]. This enhancement may qualitatively explain the large $DOS_B/DOS_A$ ratio (= 1500), since $\Delta_5$ and $\Delta_2$ bands involve hybridized states with notable $d$-orbital characters [39].

Finally, we propose a possible physical explanation for the emergence of highly spin-polarized Path-A, despite the presence of polycrystalline Fe and amorphous MgO/MgAl$_2$O$_4$. First, we expect that the tunneling current is primarily contributed by Fe electron states near $k_\parallel = 0$, for two reasons: (i) the radius of Si 2D Fermi sphere along the [001] the conduction-band valleys is much smaller than that of Fe, suppressing tunneling at larger $k_\parallel$. (ii) The conduction band minima $E_C$ of both crystalline MgO and MgAl$_2$O$_4$ are located at the $\Gamma$ point, and $E_C$ is expected to remain near the $\Gamma$ point in their amorphous phases. Second, first-principles calculations show that the $E_C$ of MgO and MgAl$_2$O$_4$ consists of the band with $\Delta_1$ symmetry, namely, the identity representation [39,40,42]. This characteristic may promote the tunneling of Fe bands with matched symmetries (e.g. $\Delta_1$, $\Sigma_1$, and $\Lambda_1$), even in the presence of structural disorder in the polycrystalline phase. The calculations clearly show the notable feature that these symmetry-matched bands are highly spin-polarized and exhibits s-like dispersion [47], which could account for the experimentally observed high spin polarization of Path-A ($P_{S,A} = 32\%$).

|  | MgO/MgAl$_2$O$_4$ | MgO [42] | MgO [40] | MgAl$_2$O$_4$[42] |
|---|---|---|---|---|
| $\lambda_A$ (nm) | 0.44 | | | |
| $\lambda_B$ (nm) | 0.1 | | | |
| $\lambda_{\Delta 1}$ (nm) | | 0.16 | 0.24 | 0.30 |
| $\lambda_{\Delta 5}$ (nm) | | 0.067 | 0.063 | 0.13 |
| $\lambda_A/\lambda_B$ (or $\lambda_{\Delta 1}/\lambda_{\Delta 5}$) | 4.4 | 2.4 | 3.8 | 2.3 |

Table 2 Experimentally extracted fitting parameters $\lambda_A$ and $\lambda_B$ and theoretically calculated $\lambda_{\Delta 1}$ and $\lambda_{\Delta 5}$ values at $k_\parallel = 0$ in the 2D Brillouin zone of the Fe (001) plane for MgO and MgAl$_2$O$_4$ barriers. The last row lists the calculated ratios $\lambda_A/\lambda_B$ and $\lambda_{\Delta 1}/\lambda_{\Delta 5}$.

**VII-D Other possible physical origins for Path-B**

In Sec. VII-C, we discussed that Fe bands with notable $d$-orbital characters may offer an intrinsic explanation for the emergence of Path-B. In particular, Fe $d$-band surface states could account for the exceptionally large $DOS_B/DOS_A$ ratio (= 1500). However, given the magnitude of this ratio, extrinsic mechanisms cannot be completely ruled out in our junctions. Among these, interfacial effects are considered the most likely contributors. This inference is supported by the analysis in Secs. VII-A and VII-B, which allowed us to exclude magnetically dead layers (MDLs) and hopping conduction via barrier defects as primary causes of the observed enhancement of $DOS_B/DOS_A$. It is well recognized that interface-related imperfections, such as defects, strain, and structural disorder, can generate localized states that enhance the local density of states and thereby increase the $DOS_B/DOS_A$ ratio.

It is worth noting that these extrinsic mechanisms cannot be conclusively confirmed and ruled out using the current characterization techniques, such as electrical measurements or TEM images. Despite these, the present work has significantly narrowed the range of plausible physical origins of Path-B. Further clarifications are still needed with more advanced experimental and theoretical investigations on interfacial effects.

## VIII. Conclusion

We systematically studied the material properties and spin transport mechanisms in original polycrystalline Fe/amorphous-MgO($t_{MgO}$)/amorphous-MgAl$_2$O$_4$($t_{MAO}$)/$n^+$-Si junctions, revealing the distinctive dependence of spin polarization on the barrier thickness across an extensive range of $t_{MgO}$ and $t_{MAO}$ combinations. This dependence arises from two key material advantages of our junctions (i) the atomically sharp MgAl$_2$O$_4$/$n^+$-Si interface enables the detection of the spin signal with the thin tunnel barrier layer, and (ii) the Fe thin film in contact with MgO exhibits good magnetic properties for the enhanced spin signal amplitudes in the thicker tunnel barrier range. Thus, the Fe/MgO interface is essential for efficient spin injection, while the use of MgO on the Si side is not critical. Furthermore, the observed dependence appears to be general across various ferromagnetic metal (FM)/insulator(I)/Si junctions: Data points extracted from previous studies closely follow the rise in spin polarization as the barrier thickness is increased. This feature is well explained by our phenomenological "two-path" model that incorporates a highly spin-polarized Path-A and weakly spin-polarized Path-B. Therefore, the combination of our experimental results and theoretical model provides a unified framework for understanding the spin transport physics in FM/I/Si structures.

Applications of semiconductor (SC)-based spin transport devices require ferromagnetic tunnel junctions that achieve both high spin injection/detection efficiency and sufficiently low junction resistance. Although our experimental results indicate that achieving such ideal junction characteristics is challenging, as far as the polycrystalline-FM/amorphous-I/SC structure is used, the physical insights gained from our model offer valuable guidance to overcome the difficulties through the band structure and material engineering of FM/I/SC structures.

**Acknowledgments**
This work was partly supported by Grants-in-Aid for Scientific Research (20H05650, 23K17324, 23H00177, 25H00840), CREST Program (JPMJCR1777) of Japan Science and Technology Agency, and Spintronics Research Network of Japan (Spin-RNJ). B.Y. thanks the financial support from the MERIT-WINGS Program at the University of Tokyo.

**Data availability**
The data that support the findings of this article are openly available [48].

**Figures**

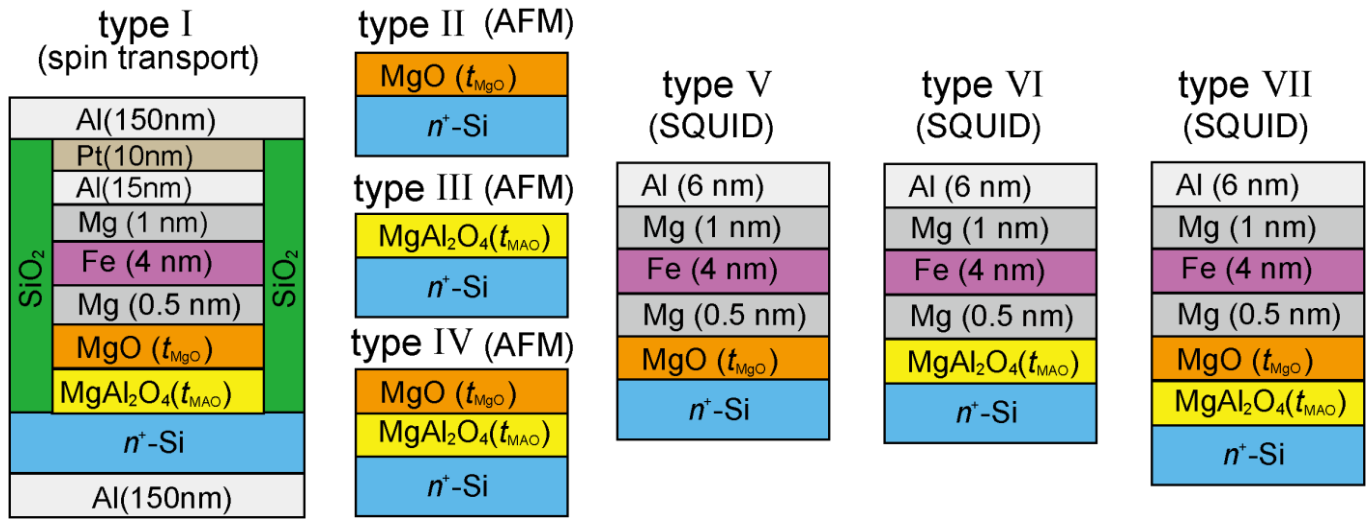

FIG.1. Layered structures examined in this study: type I was used for electrical and spin tunneling measurements and transmission electron microscopy (TEM), type II−IV were used for atomic force microscopy (AFM), and type V−VII for superconducting quantum interference device (SQUID) magnetometry. The naming convention is used consistently throughout this work.

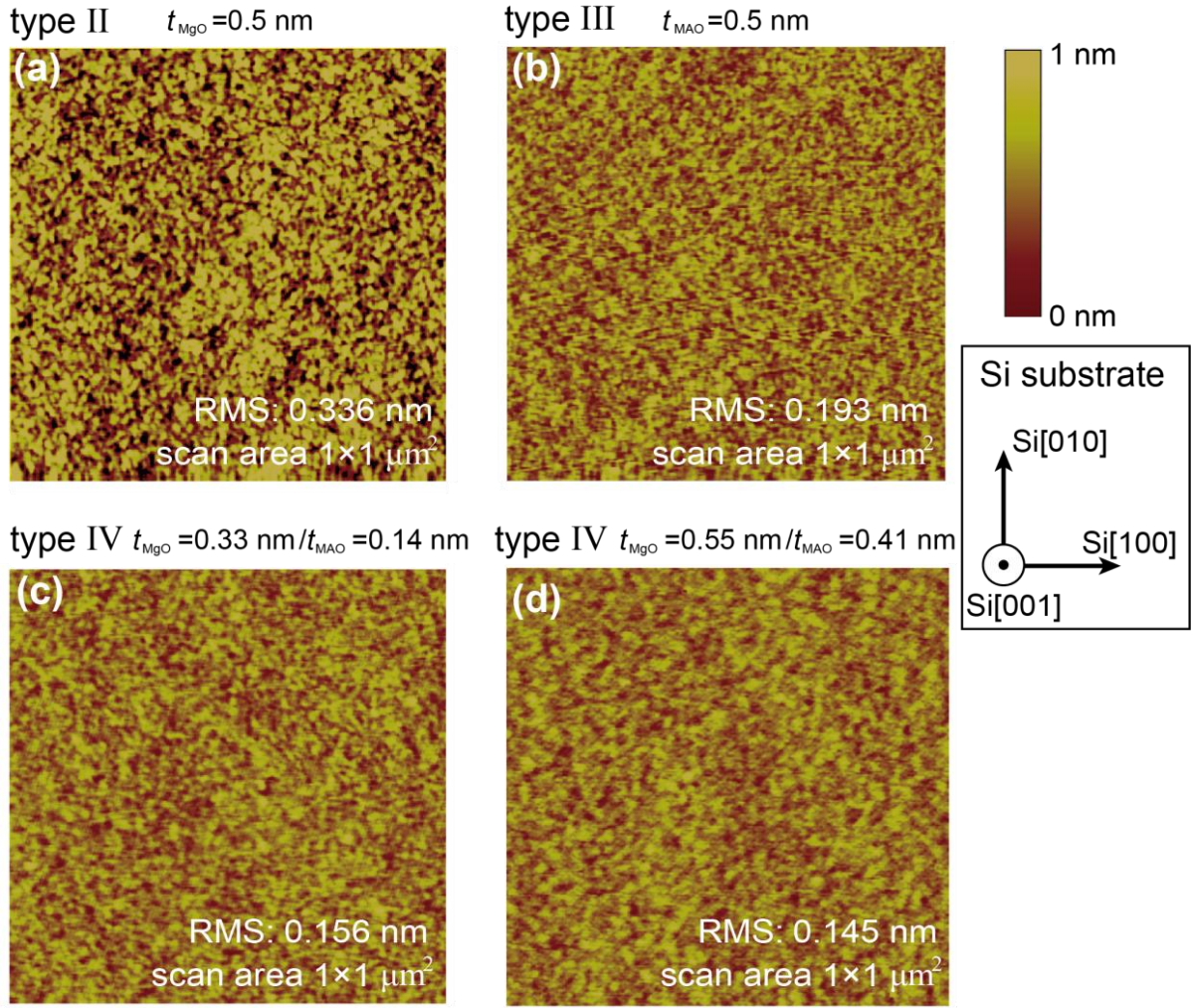

**FIG. 2.** (a) an AFM image of a type II structure with $t_{MgO} = 0.5$ nm. (b) an AFM image of a type III structure with $t_{MAO} = 0.5$ nm. (c, d) AFM images of a type IV structure with MgO/MgAl$_2$O$_4$ thickness combinations of $t_{MgO} = 0.33$ nm/$t_{MAO} = 0.14$ nm and $t_{MgO} = 0.55$ nm/$t_{MAO} = 0.41$ nm, respectively. The estimated RMS values are 0.336, 0.193, 0.156, and 0.145 nm for (a−d), respectively. All images were obtained over a 1×1 μm² scan area. The crystallographic directions of the Si substrate are indicated on the right side of the figure.

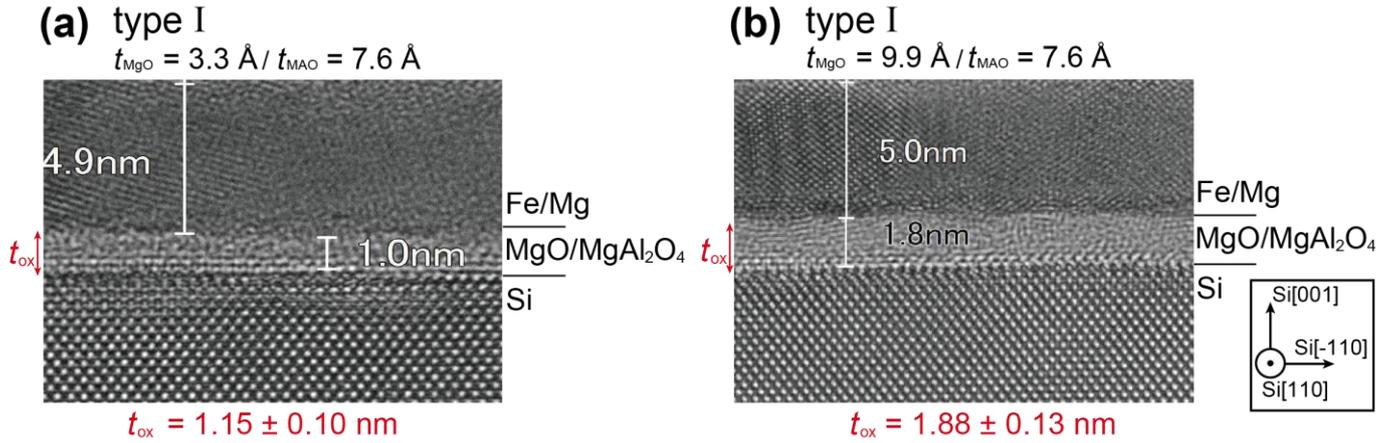

**FIG. 3.** Cross-sectional transmission electron microscopy (TEM) image of a type I structure with MgO/MgAl$_2$O$_4$ combinations of (a) $t_{MgO}$ = 3.3 Å/$t_{MAO}$ = 7.6 Å and (b) $t_{MgO}$ = 9.9 Å/$t_{MAO}$ = 7.6 Å, where the incident electron beam was aligned along the [110] crystallographic axis of Si. The total oxide thicknesses were estimated to be $t_{ox}$ = 1.15 ± 0.10 nm in (a) and $t_{ox}$ = 1.88 ± 0.13 nm in (b), respectively.

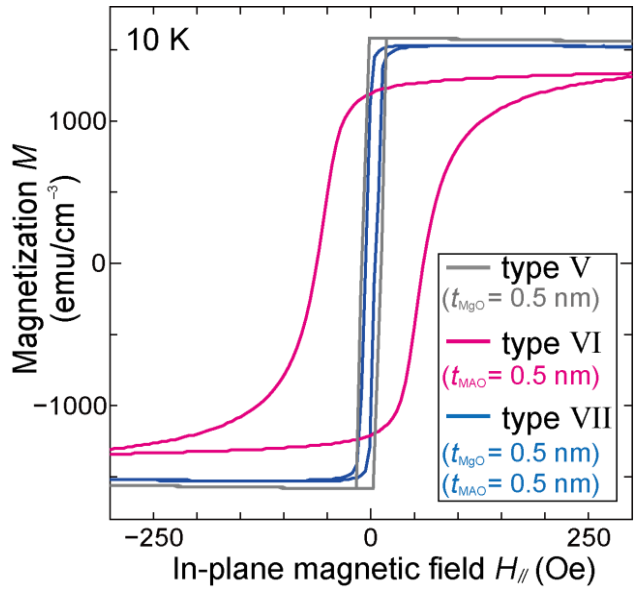

**FIG. 4.** Magnetization versus in-plane magnetic field ($M$–$H_{//}$) curves measured by SQUID magnetometry at 10 K for type V ($t_{MgO}$ = 0.5 nm), type VI ($t_{MAO}$ = 0.5 nm), and type VII ($t_{MAO}$ = 0.5 nm/$t_{MgO}$ = 0.5 nm), corresponding to the gray, blue, and pink curves, respectively.

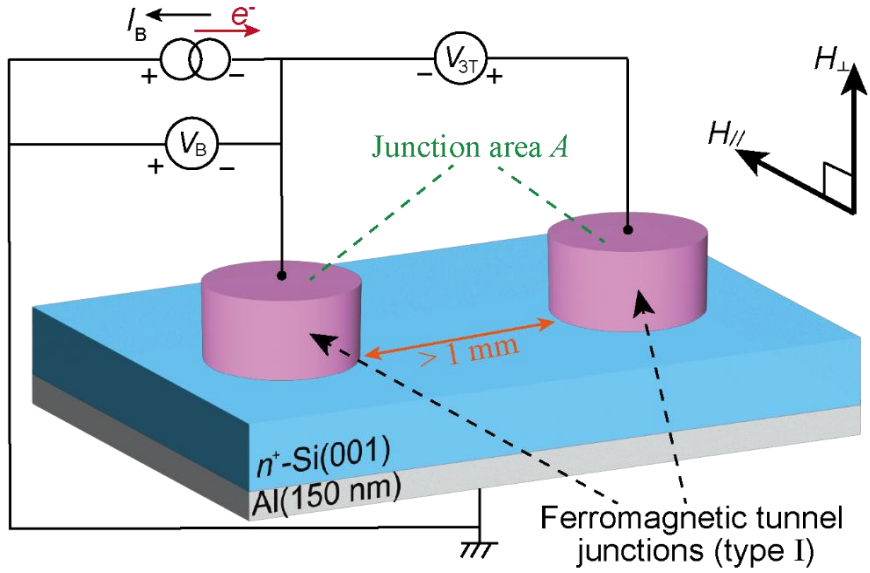

**FIG. 5.** Schematic illustration of the device with a type I structure and three-terminal Hanle (3TH) measurement setup, where the two junctions are electrically isolated by the thermally oxidized SiO$_2$ layer (not shown). A positive (negative) $I_B$ corresponds to the spin injection (spin extraction) geometry. The in-plane magnetic field $H_{//}$ and perpendicular-to-plane magnetic field $H_\perp$ are defined along the Si [001] and Si[100] directions, respectively.

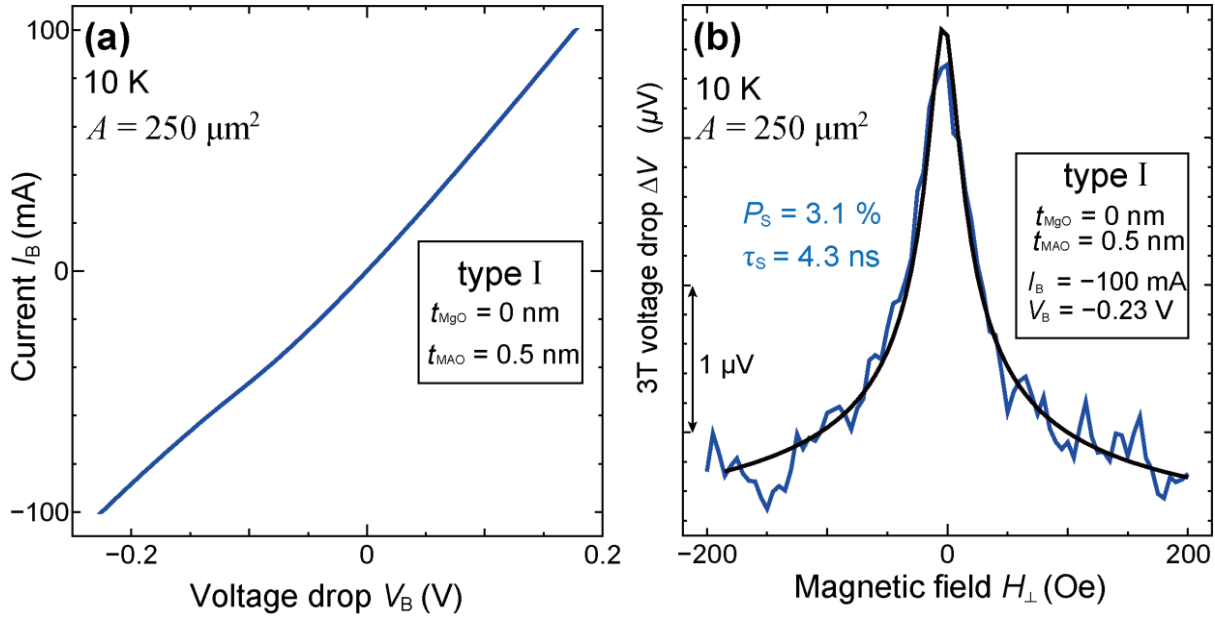

**FIG. 6** (a) I-V characteristics and (b) N-3TH signal (blue) measured at 10 K for the device with the type I structure, where $t_{MAO}$ = 0.5 nm and $t_{MgO}$ = 0 nm, and a perpendicular-to-plane magnetic field $H_\perp$ was swept from −200 to 200 Oe. The junction area is $A$ = 250 μm$^2$, and N-3TH signal was measured under $I_B$ = −100 mA. The black curve in (b) is the fitting result, from which $P_S$ =3.1% and $\tau_S$ = 4.3 ns were estimated.

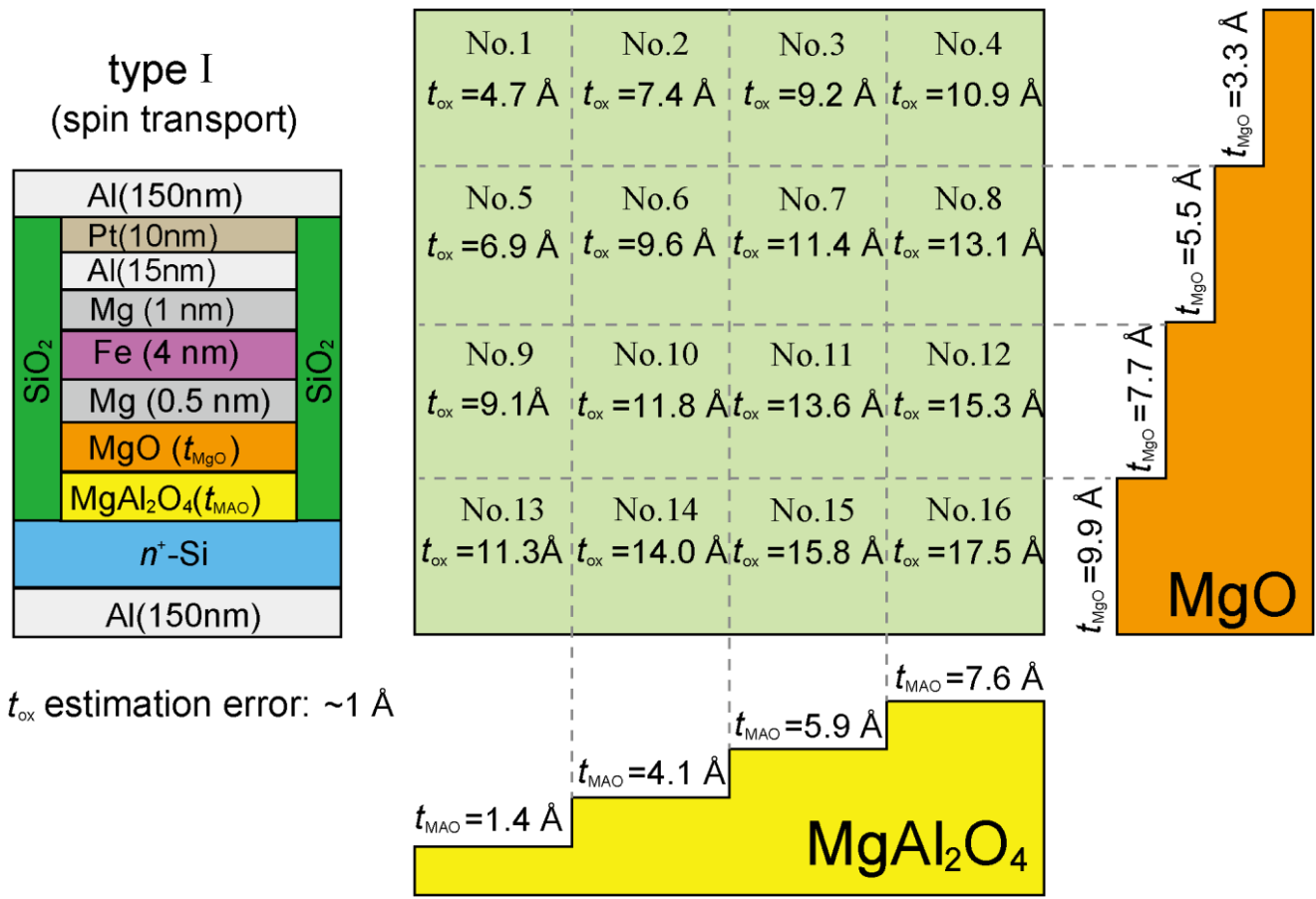

**FIG. 7.** Schematical layout of the 16 combinations of $t_{MAO}$ and $t_{MgO}$ examined for the type I structure, where $t_{MAO}$ = 1.4, 4.1, 5.9, and 7.6 Å, and $t_{MgO}$ = 3.3, 5.5, 7.7, and 9.9 Å. The typical error margin of $t_{ox}$ is approximately 1 Å.

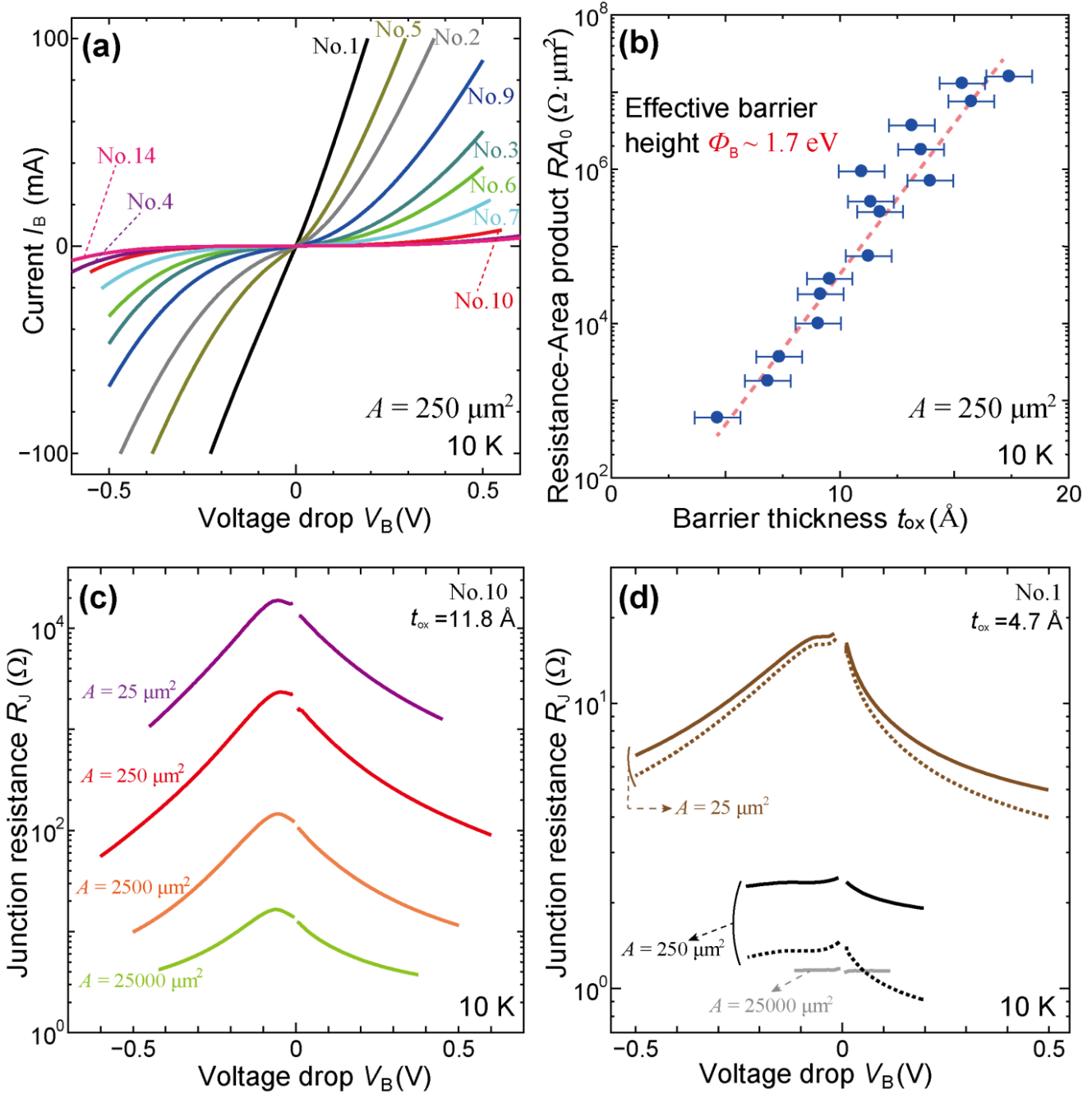

**FIG. 8.** (a) *I-V* curves measured at 10 K for No.1−7, No. 9, No.10, and No.14 combinations with $A = 250$ μm$^2$, where the correspondence between the numbers and data is indicated by color. (b) $RA_0$ values measured at 10 K with $A = 250$ μm$^2$ plotted as a function of $t_{ox}$ for all 16 combinations (No. 1−16). The dashed red curve shows a linear fitting curve, from which the effective barrier height was estimated to be $\Phi_B \sim 1.7$ eV. (c) Junction resistance $R_J$ plotted as a function of $V_B$ for No.10 combination with various junction areas at 10 K. The purple, red, orange, and green curves correspond to $A = 25, 250, 2500,$ and $25000$ μm$^2$, respectively. (d) $R_J$ versus $V_B$ for No.1 combination with $A = 25$ (black), 250 (brown), and 25000 (gray) μm$^2$ at 10 K. The dashed brown and black curves are obtained by subtracting by 1 Ω from the corresponding solid curves.

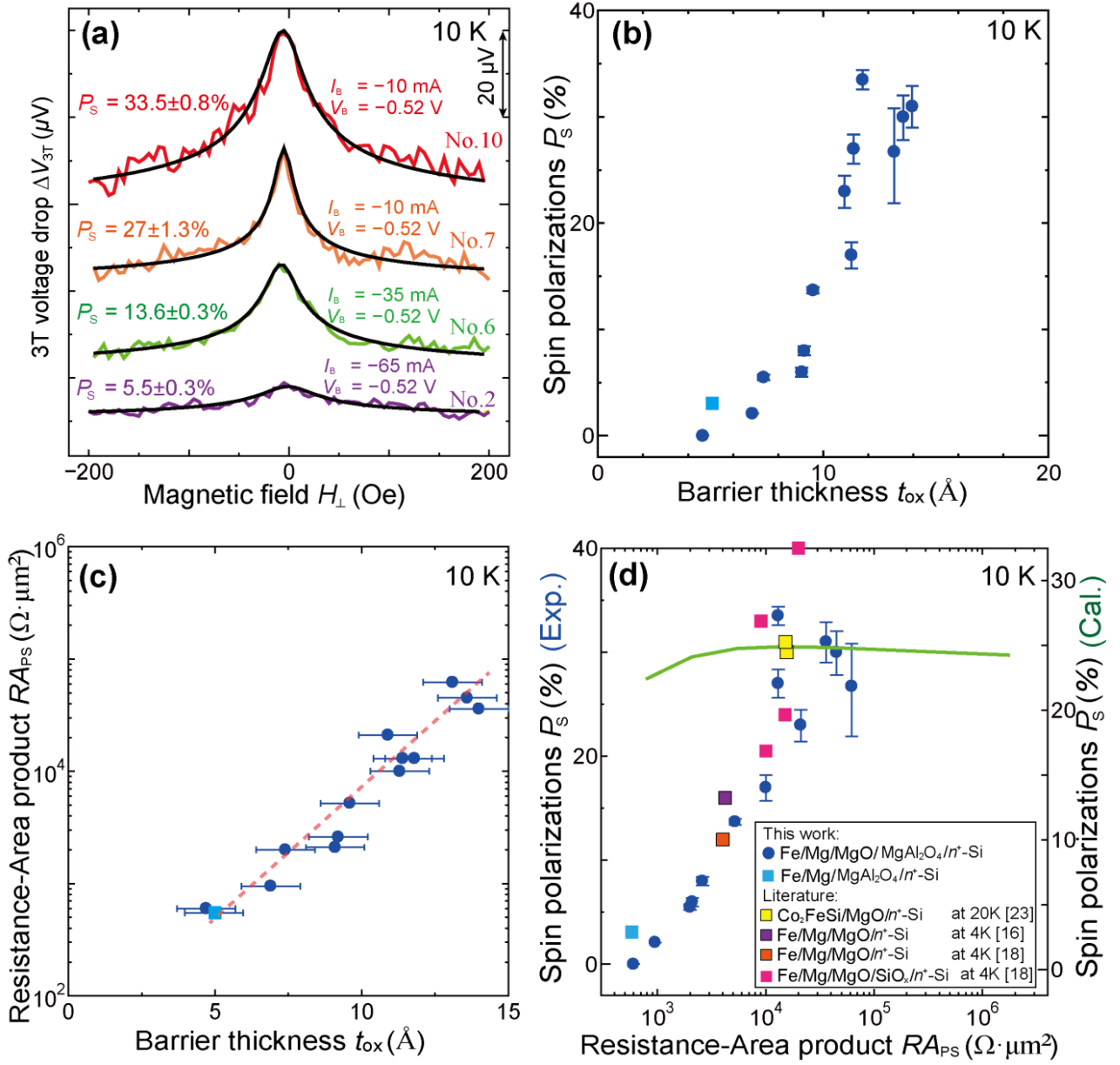

**FIG. 9.** (a) Typical N-3TH signals measured at 10 K with $A = 250$ μm$^2$ in the spin extraction geometry ($I_B < 0$) under a perpendicular magnetic field $H_\perp$ sweeping from −200 to 200 Oe. For each combination, $I_B$ was adjusted so that $V_B$ keeps a nearly constant value of −0.52 V. The red, orange, green, and purple curves represent the experimental $\Delta V_{3T}$ signals measured for No.10, No.7, No.6, and No.2 combinations, respectively, while the black curves are the corresponding fitting results obtained using Eq. (1). (b) Spin polarization $P_S$ plotted as a function of $t_{ox}$ at 10 K for various $t_{MAO}/t_{MgO}$ combinations (shown in Fig. 7), where data for No. 12, No. 15, and No. 16 combinations are excluded due to large error margins. The blue square is extracted from Figs. 6(a) and (b) for the Fe/MgAl$_2$O$_4$/$n^+$-Si junction. (c) $RA_{PS}$ (= $A \times V_B / I_B$) values measured at 10 K plotted as a function of $t_{ox}$, where the dashed red curve shows a linear fitting curve. The blue square is extracted from Figs. 6(a) and (b) for the Fe/MgAl$_2$O$_4$/$n^+$-Si junction. (d) Blue dots: experimentally estimated $P_S$ plotted as a function of $RA_{PS}$ at 10 K. Blue square: the data point extracted from the Fe/MgAl$_2$O$_4$(0.5 nm)/$n^+$-Si junction in Figs. 6(a) and (b). Green dashed curve: a theoretical $P_S - RA_{PS}$ curve calculated from Eq. (2). Colored squares: experimental ($P_S$, $RA_{PS}$) data points extracted from Ref. [16,18,23].

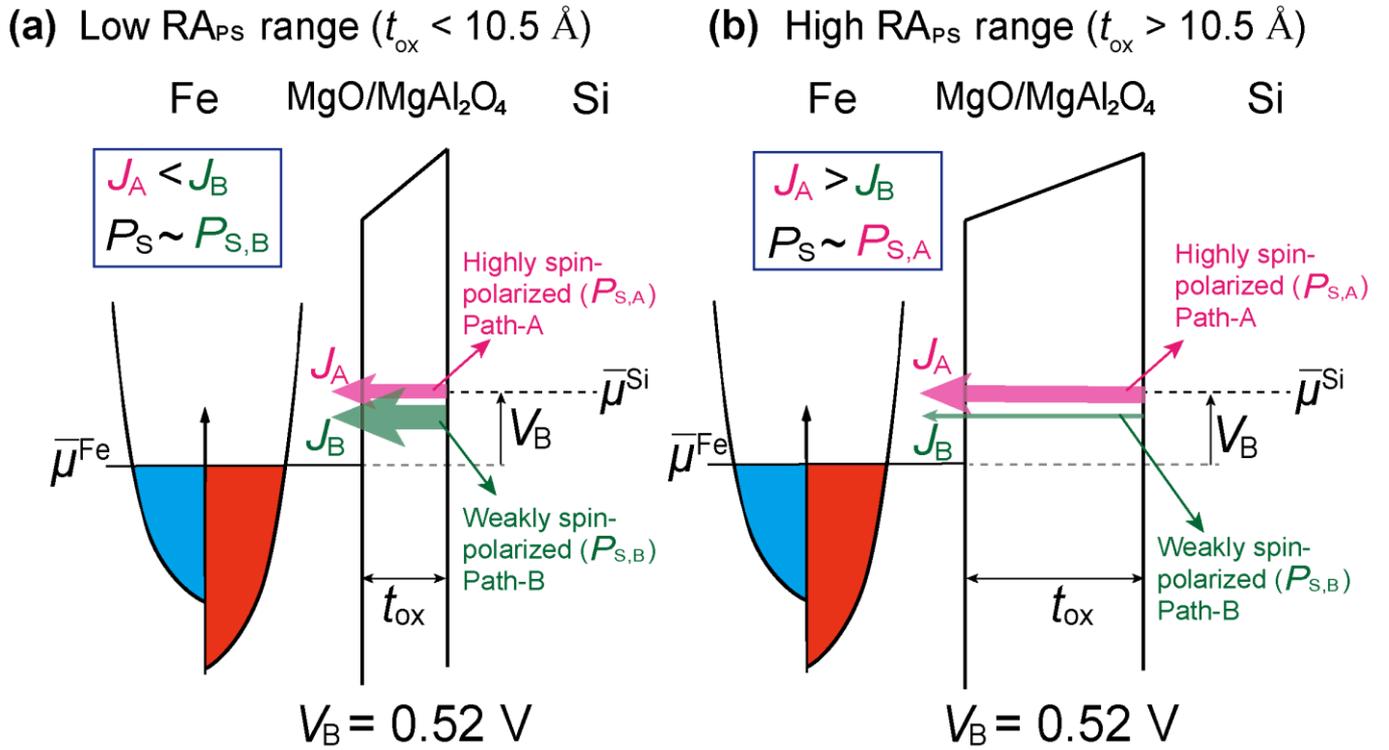

**FIG. 10.** Schematic band diagrams of spin transport process at a constant $V_B$ (= 0.52 V) bias in the low (a) and high (b) $RA_{PS}$ ranges, respectively, where $\bar{\mu}^{Fe}$ and $\bar{\mu}^{Si}$ are the Fermi levels of Fe and $n^+$-Si, respectively, red and blue colors in Fe represent the filled states of up-spin and down-spin electrons, respectively, $J_A$ and $J_B$ are the electron current densities of Path-A and Path-B, and $P_{S,A}$ and $P_{S,B}$ are the electron spin polarizations of Path-A and Path-B, respectively, with $P_{S,A}$ being larger than $P_{S,B}$. The widths of the pink and green arrows express the magnitudes of $J_A$ and $J_B$, respectively.

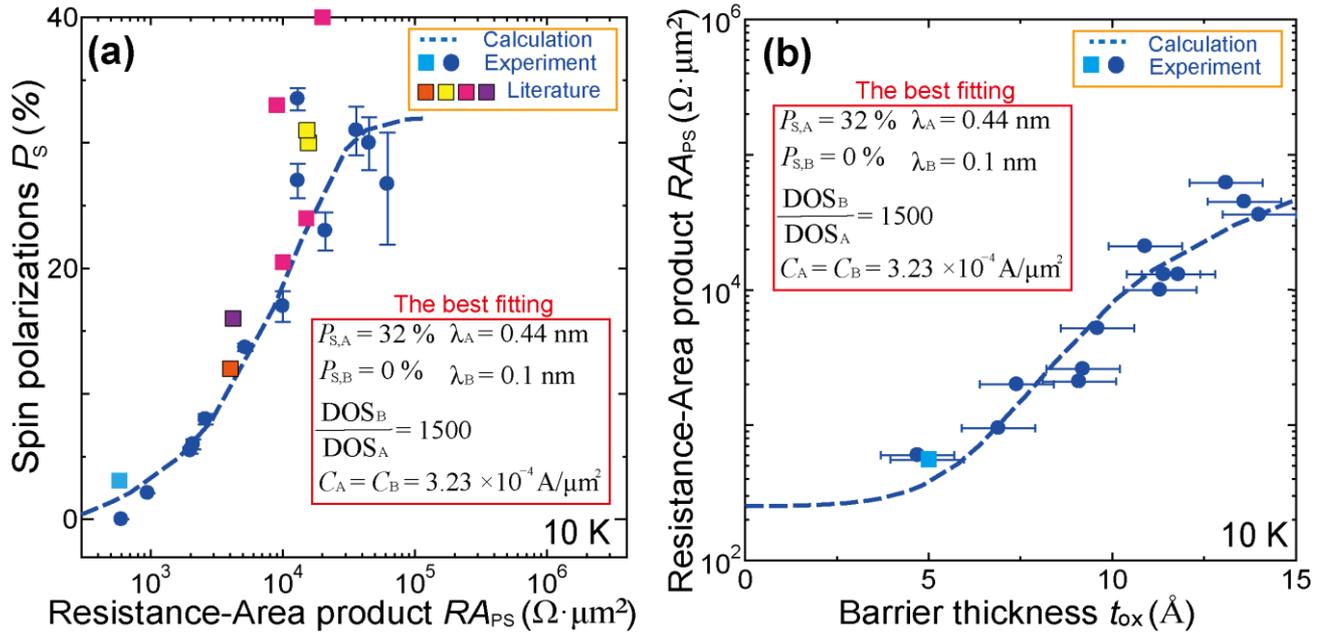

**FIG. 11.** Experimental (a) $P_S - RA_{PS}$ and (b) $RA_{PS} - t_{ox}$ plots (blue dots), extracted from Figs. 9(d) and (c), respectively. The blue dashed curves represent the best-fitting results calculated using Eqs. (3) and (4), where the fittings parameters are listed in the red rectangle.

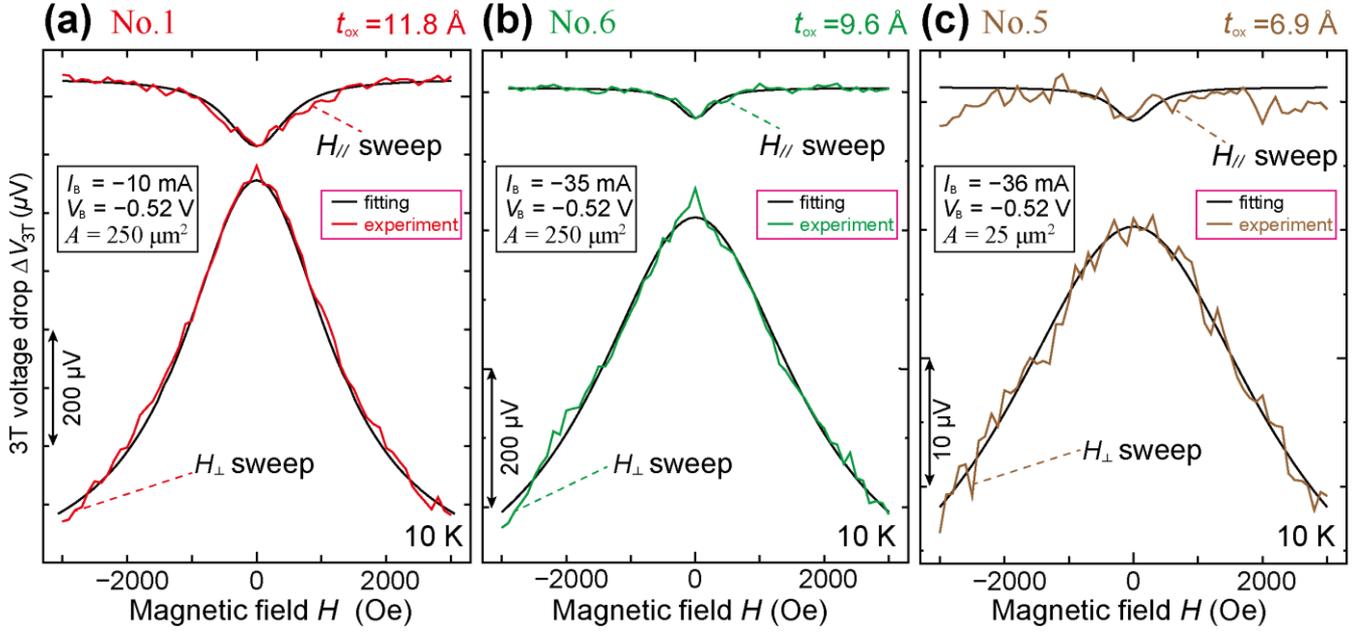

**FIG.12.** (a)-(c) Experimental I-3TH (top) and B-3TH (bottom) signals measured at 10 K for No.10 (red), No.6 (green), and No. 5 (brown) combinations, with $A$ = 250, 250, and 25 µm², respectively, where an in-plane magnetic field $H_{//}$ and a perpendicular-to-plane magnetic field $H_\perp$ were swept from −3000 to 3000 Oe. The black curves represent the corresponding Lorentzian fitting results. For each combination, $I_B$ was adjusted so that $V_B$ keeps a nearly constant value of −0.52 V.

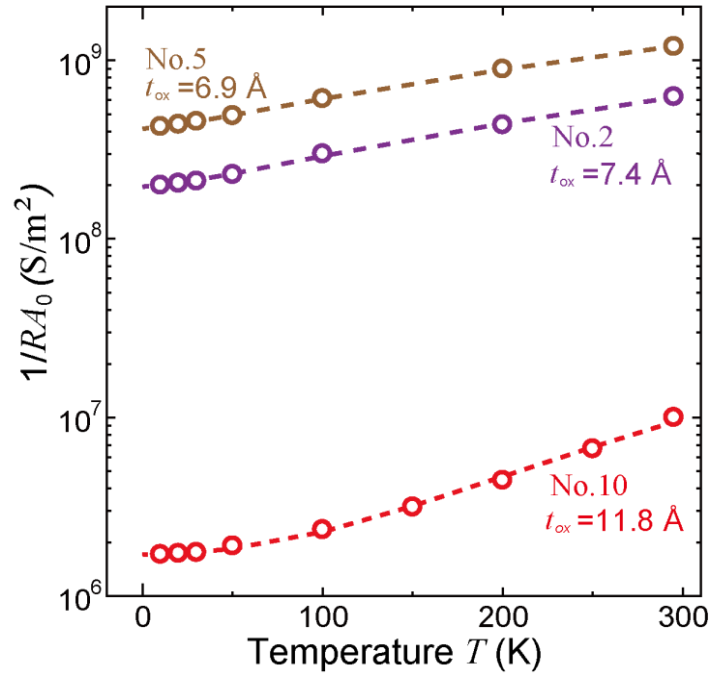

**FIG. 13.** Temperature dependence of the junction conductance per area (open dots), $\sigma_0 = 1/RA_0$, for No.10 (red), No.2 (purple), and No.5 (brown) combinations, respectively. For each combination, the dashed curves in the corresponding colors represent their fittings calculated using Eq. (10).

# Supplemental Material

# Spin injection in Si-based ferromagnetic tunnel junctions with MgO/MgAl$_2$O$_4$ barriers: Experimental and theoretical investigation of barrier thickness-dependent spin tunneling efficiency

Baisen Yu[1], Shoichi Sato[1, 2], Masaaki Tanaka[1, 2, 3], and Ryosho Nakane[1, 3, 4]

[1]*Deptartment of Electrical Engineering and Information Systems, The University of Tokyo,
7-3-1 Hongo, Bunkyo-ku, Tokyo 113-8656, Japan*
[2]*Center for Spintronics Research Network (CSRN), The University of Tokyo,
7-3-1 Hongo, Bunkyo-ku, Tokyo 113-8656, Japan*
[3]*Institute for Nano Quantum Information Electronics, The University of Tokyo,
4-6-1 Meguro-ku, Tokyo 153-8505, Japan*
[4]*Systems Design Lab (d.lab), The University of Tokyo,
7-3-1 Hongo, Bunkyo-ku, Tokyo 113-8656, Japan*

**S1. Error estimation in the 3TH signals based on the root-mean-square deviation (RMSD) method**

Here we describe how the error of the spin polarization $P_S$ was estimated by fitting Eq. (1) to the 3TH signals, using No.10 combination as an example, as shown in Fig. S1. The experimental red curve and black fitting curve are extracted from those in Fig. 9(a). From this fitting, the amplitude of the N-3TH signal was estimated to be $\Delta V_{3T}$ = 42.6 µV, and the corresponding spin polarization was estimated to be $P_S$ =33.5%. The estimation error of $\Delta V_{3T}$ was calculated using the root-mean-square deviation (RMSD), defined as:

$$\Delta V_{3T,RMSD} = \sqrt{\frac{1}{n}\sum_{i=1}^{n}\left(\Delta V_{3T,exp}^i - \Delta V_{3T,cal}^i\right)^2}, \quad (S1)$$

where $\Delta V_{3T,exp}^i$ and $\Delta V_{3T,cal}^i$ are the experimental and calculated amplitudes of the N-3TH signals at each applied perpendicular magnetic field $H_\perp$, respectively, and $n$ is the total number of data points. Using this approach, we obtained $\Delta V_{3T,RMSD}$ = 2.4 µV. The upper and lower bounds of $\Delta V_{3T}$ are given by

$$\Delta V_{3T,min} = \Delta V_{3T} - \Delta V_{3T,RMSD} = 40.2 \text{ µV},$$

$$\Delta V_{3T,max} = \Delta V_{3T} + \Delta V_{3T,RMSD} = 45.0 \text{ µV},$$

respectively. The corresponding lower and upper errors for $P_S$ were then calculated from $\Delta V_{3T,min}$ and $\Delta V_{3T,max}$ values, which were estimated to be 32.7% and 34.3%, respectively. In this manner, we calculated the errors for other combinations (blue dots) shown in Figs. 9(b) and (d).

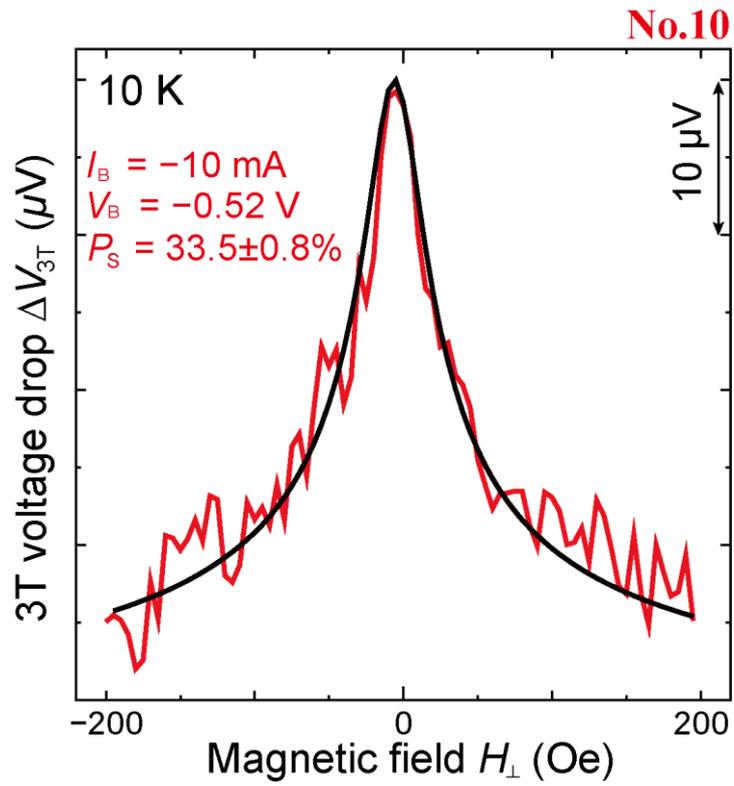

**FIG. S1.** N-3TH signal (red) measured at 10 K for No.10 combination and the corresponding fitting curve (black), which are extracted from Fig. 9(a).

## S2. 3TH signals in a Fe/MgAl$_2$O$_4$(0.5 nm)/$n^+$-Si junction with various $I_B$ values

In Sec. IV-A, we showed the experimental 3TH signal for the Fe/MgAl$_2$O$_4$(0.5 nm)/$n^+$-Si structure measured at $I_B = -100$ mA and $V_B = -0.23$ V at 10K. To confirm the applicability of this structure as a spin injector/detector, here we present 3TH signals and corresponding fitting curves for the same device with different $I_B$ values. Figure S2 shows the experimental 3TH signals measured at 10 K with $A = 250$ μm$^2$. The blue, light blue, green, and pink curves correspond to the measurements with $I_B = -100, -70, -50$, and $-30$ mA, respectively. The black curves show the corresponding fitting curves using Eq. (1). The spin polarization and spin lifetime were estimated to be $P_S = 3.1, 3.1, 3.2$, and 1.8% and $\tau_S = 4.3, 2.9, 3.3$, and 3.3 ns, respectively. The comparable $\tau_S$ values confirm that these signals originate from spin injection into $n^+$-Si. The $V_B$ dependence of $P_S$ is qualitatively consistent with our previous model [S1]: The slight increase in $P_S$ from 1.8% to 3.1% is attributed to a steep increase of $P_{det}$ as $V_B$ is negatively increased from 0, while the nearly constant $P_S$ at higher $|V_B|$ (from $-0.11$ to $-0.23$ V) reflects that both $P_{inj}$ and $P_{det}$ undergo modest changes under such small variations in $V_B$.

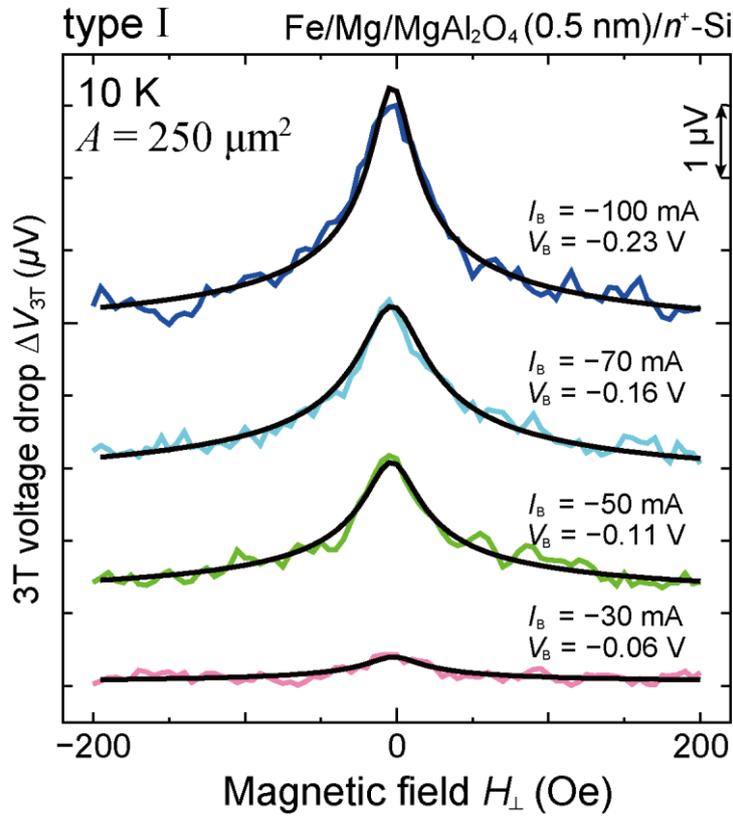

**FIG. S2.** N-3TH signals measured at 10 K with $A = 250$ μm$^2$ in the spin extraction geometry ($I_B < 0$). The blue, light blue, green, and pink curves represent the experimental $\Delta V_{3T}$ signals measured with $I_B = -100, -70, -50$, and $-30$ mA, respectively. The black curves show the corresponding fitting curves using Eq. (1).

## S3. 3TH signals in Fe/MgAl$_2$O$_4$/$n^+$-Si tunnel junctions with different barrier thickness

In Secs. IV-A and S2, we demonstrated that the type I structure with $t_{MAO} = 0.5$ nm and $t_{MgO} = 0$, Fe(4 nm)/Mg(0.5 nm)/MgAl$_2$O$_4$(0.5 nm)/$n^+$-Si, is applicable to a spin injector and detector, with typical spin polarization $P_S = 3\%$ and spin lifetime $\tau_S = 3$ ns, respectively. However, clear 3TH signals were not observed when $t_{MAO}$ is further increased. As shown in Fig. S3(a), the blue and orange curves represent the $I$-$V$ curves measured at 10 K with $A = 250$ μm$^2$ for the Fe(4 nm)/Mg(0.5 nm)/MgAl$_2$O$_4$($t_{MAO}$)/$n^+$-Si structure with $t_{MAO} = 0.5$ and 0.9 nm, respectively (the blue curve is extracted from Fig. 6(a)). The stronger nonlinearity of the orange curve reflects the increased $t_{MAO}$ value. Figure S3(b) shows the corresponding 3TH signals measured at 10 K in the spin extraction geometry ($I_B < 0$) under a perpendicular magnetic field $H_\perp$, where the blue curve is extracted from Fig. 6(b). Due to the increased barrier thickness and enhanced $V_B$ value for the orange curve, the SNR deteriorated, and the noise level reached approximately 1 μV, which is comparable to the expected amplitude of 3TH signals at $I_B = -20$ mA when $P_S = 3\%$ is assumed. As a result, convincing 3TH signals were not observed. It is noted that 3TH signals were observed at 4 K in Fe/MgO/$n^+$-Si junctions with a similar MgO barrier thickness of 0.8 nm [S2]. These results indicate that, unlike the MgO-based tunnel junctions, the Fe/MgAl$_2$O$_4$($t_{MAO}$)/$n^+$-Si structure with moderately thick MgAl$_2$O$_4$ layer is not suitable for achieving highly efficient spin injection. As discussed in Sec. IV-A, this phenomenon may be associated with the deteriorated Fe/MgAl$_2$O$_4$ interface, on which the Fe thin film shows the degraded $M_S$ and $M_r$ values. Thus, we used the type I structure with nonzero $t_{MAO}$ and nonzero $t_{MgO}$ values to perform 3TH measurements and investigate the dependence of $P_S$ on the barrier thickness.

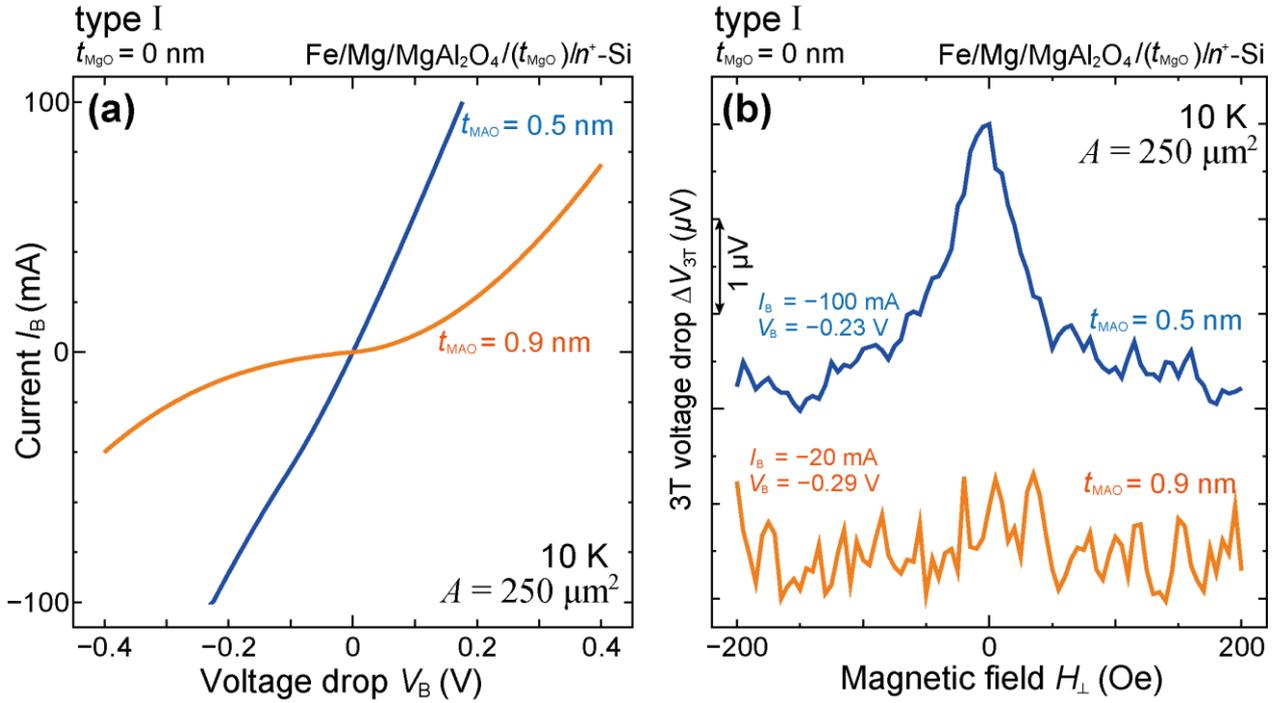

**FIG. S3.** (a) $I$-$V$ characteristics and (b) 3TH signals measured at 10 K for type I structure and $A = 250$ μm$^2$, where a perpendicular-to-plane magnetic field $H_\perp$ was swept from −200 to 200 Oe in (b). The orange and blue curves were measured for the Fe/MgAl$_2$O$_4$($t_{MAO}$)/$n^+$-Si junctions with $t_{MAO} = 0.5$ and 0.9 nm, respectively.

## S4. Variations of $P_S$ with small deviations from $V_B = -0.52$V in Fe/MgO/MgAl$_2$O$_4$/$n^+$-Si tunnel junctions

In Sec. IV-C, we selected $V_B = -0.52$ V as a constant bias condition to measure the N-3TH signals and estimate the $P_S$ values for the Fe/MgO/MgAl$_2$O$_4$/$n^+$-Si structure with various $t_{MAO}$ and $t_{MgO}$ combinations. To confirm that the $P_S$ values estimated under this bias are representative for each combination, we experimentally examined the fluctuations of N-3TH signals and estimated $P_S$ values for No.6, No. 7, and No. 9 combinations under slight variations in $V_B$ around $-0.52$ V. Figure S4 shows the 3TH signals measured for No.7 with $A = 250$ μm$^2$ at 10 K under a swept perpendicular-to-plane magnetic field $H_\perp$. The pink, orange, and purple curves correspond to experimental $\Delta V_{3T}$ signals measured with $I_B = -5, -10,$ and $-15$ mA, respectively, with corresponding $V_B$ values of $-0.44, -0.52,$ and $-0.57$ V. The orange curve is extracted from Fig. 9(a). The estimated $P_S$ values for No.7 combination are summarized in the second row of Table S1, with typical errors below 1%. These results show that $P_S$ is relatively insensitive to small deviations in $V_B$ (less can $\pm 0.1$ V) from 0.52V. A similar trend was also observed for No.6 and No. 9 combinations, as listed in Tabel S1, where $P_S$ showed very modest change as $V_B$ is negatively decreased from $-0.44$V to $-0.52$V. Therefore, the $P_S$ value obtained at $V_B = -0.52$V is representative of the spin injection/detection efficiency for each combination. In other words, our $P_S - t_{ox}$ curve obtained under $V_B = -0.52$V is characteristic and robust with respect to small variations in the bias voltage.

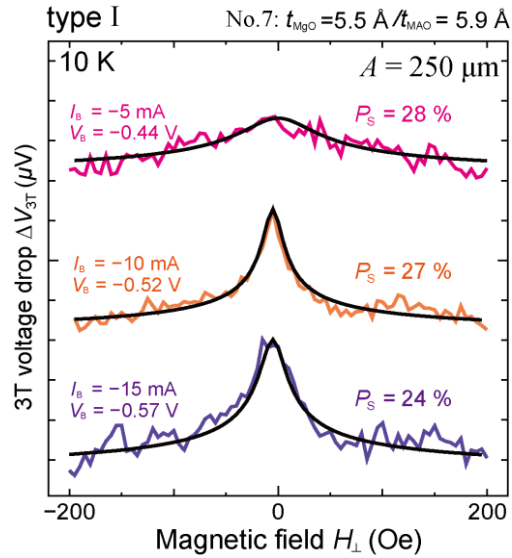

**FIG. S4.** N-3TH signals measured at 10 K for No.7 combination with $A = 250$ μm$^2$ in the spin extraction geometry ($I_B < 0$). The pink, orange, and purple curves represent the experimental $\Delta V_{3T}$ signals measured with $I_B = -5, -10,$ and $-15$ mA, respectively. The black curves show the corresponding fitting curves using Eq. (1)

|       | $V_B$ / $P_S$   | $V_B$ / $P_S$   | $V_B$ / $P_S$   |
|-------|-----------------|-----------------|-----------------|
| No.6  | 0.42V / 15.2%   | 0.48V / 13.5%   | 0.52V / 13.6%   |
| No.7  | 0.44V / 28%     | 0.52V / 27%     | 0.57V / 24%     |
| No.9  | 0.44V / 6.7%    | 0.52V / 6%      |                 |

**Table S1** Summarized $V_B$ / $P_S$ data points at 10 K for No. 6, No. 7, and No. 9 combinations with slight variations in $V_B$ around $-0.52$ V, where typical errors is below 1%.

## S5. Comparison of experimental $P_S - RA_{PS}$ characteristics with the conductivity mismatch model

Our experimental $P_S - RA_{PS}$ characteristics in Sec. V-B show a clear increase in $P_S$ as $RA_{PS}$ (or $t_{ox}$) is increased. This trend is qualitatively analogous to the well-known expression for the conductivity mismatch problem [S3]:

$$P_S \propto P_{inj} = \frac{\gamma}{1 + \frac{r_{sr,Si}}{RA_{PS}}}, \qquad (S1)$$

where $\gamma$ is the spin asymmetric factor of the tunnel barrier, and $r_{sr,Si}$ is the spin resistance of Si. Equation (S1) indicates that $P_S$ increases with $RA_{PS}$, analogous to our experimental trend. However, our analysis shows that Eq. (S1) cannot account for the observed $P_S - RA_{PS}$ characteristics due to the inherently low spin resistance of $n^+$-Si. The orange dashed curve in Fig. S5 shows the theoretical $P_S - RA_{PS}$ relationship calculated by Eq. (S1), using $r_{sr} = 30\ \Omega\cdot\mu m^2$ [S2] and assuming $\gamma = 30\%$ to match the experimental $P_S$ values fluctuating around 30% in the high $RA_{PS}$ range. The blue dots are extracted from Fig. 9(b). It is evident that the orange curve deviates significantly from the experimental $P_S - RA_{PS}$ by nearly two orders of magnitude in $RA_{PS}$. In other words, the reduction in $P_S$ resulted from the conductivity mismatch problem is negligibly small when $RA_{PS}$ exceeds $1000\ \Omega\cdot\mu m^2$. Therefore, we conclude that the conductivity mismatch is not responsible for the experimental $P_S - RA_{PS}$ characteristics.

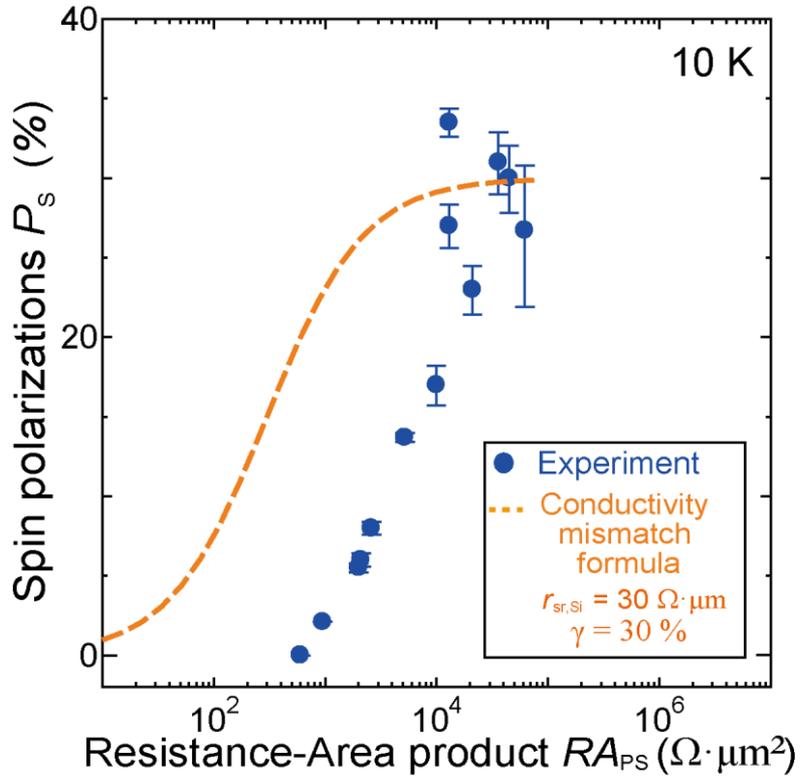

**FIG. S5.** Experimental $P_S - RA_{PS}$ characteristics (blue dots) extracted from Fig. 9(b) and calculated $P_S - RA_{PS}$ relationship using Eq. (S1) with $r_{sr} = 30\ \Omega\cdot\mu m^2$ and assuming $\gamma = 30\%$.

## S6. Numerical validation of the simplification from Eq. (2) to Eq. (3): comparison of $J_A$–$t_{ox}$ and $P_{S,A}$–$RA_{PS}$ characteristics

In Sec. VI-B, we simplified our previous model (Eq. (2)) under a constant $V_B$ into a more tractable form, Eq. (3), where $P_{S,A}$ is assumed as a constant and the exponential relationship between the current density $J_A$ and the barrier thickness $t_{ox}$ is described by the tunneling decay length $\lambda_A$. While this simplification is *not* mathematically rigorous, we assess its validity by numerically comparing the $J_A$–$t_{ox}$ and $P_{S,A}$–$RA_{PS}$ relationships obtained from Eqs. (2) and (3) separately. Figure S6(a) shows the calculated $J_A$–$t_{ox}$ relationship using Eq. (2) (green dots) by an iterative method [see details in Ref. [S1]], with $V_B = -0.52$ V and other calculation parameters adopted from Ref. [S1]. The red curve represents the $J_A$–$t_{ox}$ curve calculated using Eq. (3) with an adapted $\lambda_A = 0.1$ nm. The good agreement between these two calculations reflects that Eq. (3) provides a good approximation of Eq. (2) for reproducing the $J_A$–$t_{ox}$ behavior when an appropriate $\lambda_A$ is used. We also compared the $P_{S,A}$–$RA_{PS}$ characteristics calculated from Eqs. (2) and (3) separately, as the green dots and red curve shown in Fig. S6(b), respectively. Here, $P_{S,A} = 25\%$ is assumed in Eq. (3). It is found that the red curve (Eq. (3)) well reproduces the $P_{S,A}$–$RA_{PS}$ trend calculated by Eq. (2) when $RA_{PS} > 2000$ $\Omega \cdot \mu m^2$, while a slight deviation is observed when $RA_{PS} < 2000$ $\Omega \cdot \mu m^2$. The good agreement between the green dots and red curves confirms that Eq. (3) serves as a reliable approximation of Eq. (2) for analyzing the $P_{S,A}$–$RA_{PS}$ characteristics in our system.

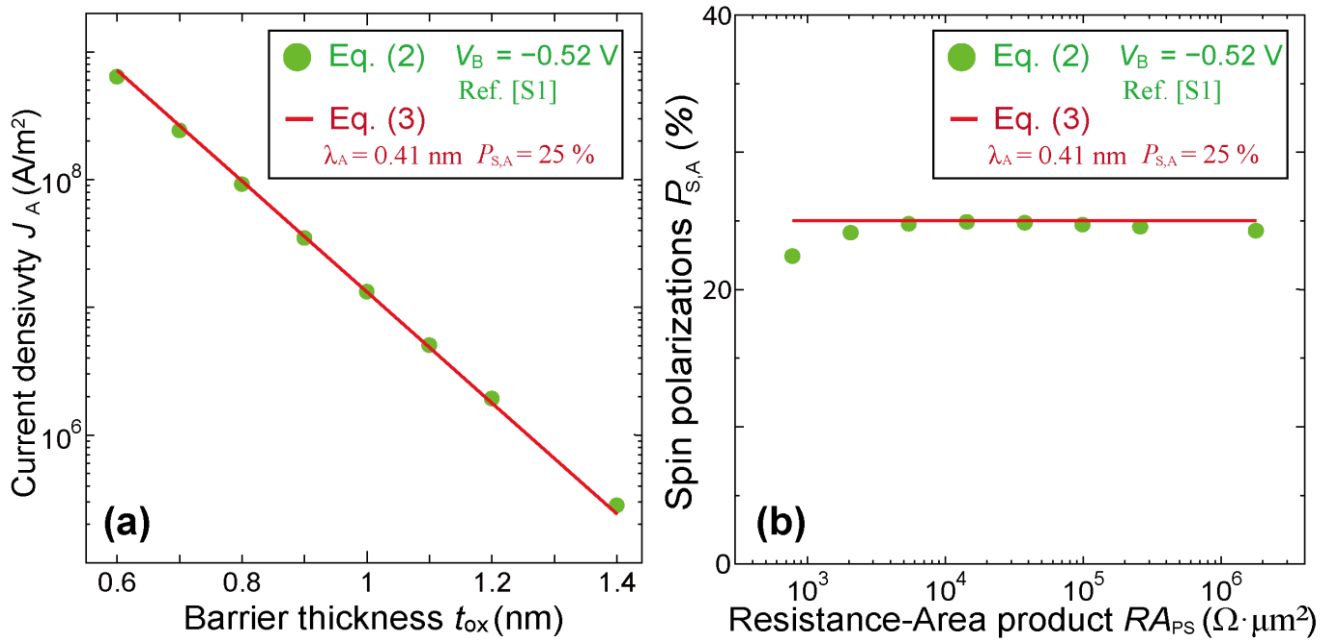

**FIG. S6.** (a) Calculated $J_A$–$t_{ox}$ characteristics using Eq. (2) (green dots) and Eq. (3) (red curve). (b) Calculated $P_{S,A}$–$RA_{PS}$ characteristics using Eq. (2) (green dots) and Eq. (3) (red curve). The calculation parameters for Eq. (2) are adopted from Ref. [S1] with $V_B = -0.52$ V, and $\lambda_A = 0.1$ nm and $P_{S,A} = 25\%$ are used for Eq. (3).

## S7. RMSD-based fitting of $P_S - RA_{PS}$ and $RA_{PS} - t_{ox}$ results

Here, we describe the procedure used to calculate the RMSD values for the fitting of the $P_S - RA_{PS}$ and $RA_{PS} - t_{ox}$ plots in Figs. 11 (a) and (b), respectively. For the $P_S - RA_{PS}$ plot, the normalized RMSD value is defined as:

$$\text{RMSD}_{\text{PS-RA}} = \sqrt{\frac{1}{n}\sum_{i=1}^{n}\left(\frac{P_{S,\text{exp}}^i - P_{S,\text{cal}}^i}{P_{S,\text{cal}}^i}\right)^2}, \tag{S2}$$

where $P_{S,\text{exp}}^i$ and $P_{S,\text{cal}}^i$ are the experimental and calculated $P_S$ values, respectively, and $n$ is the total number of data points.

For the $P_S - RA_{PS}$ plot, to account for the exponential nature of the data, the normalized RMSD value is defined using the logarithmic scale as:

$$\text{RMSD}_{\text{RA-tox}} = \sqrt{\frac{1}{n}\sum_{i=1}^{n}\left(\frac{\log_{10} RA_{PS,\text{exp}}^i - \log_{10} RA_{PS,\text{cal}}^i}{\log_{10} RA_{PS,\text{cal}}^i}\right)^2}, \tag{S3}$$

where $RA_{PS,\text{exp}}^i$ and $RA_{PS,\text{cal}}^i$ are the experimental and calculated $RA_{PS}$ values, respectively, and $n$ is the total number of data points. The total RMSD is defined as the sum of the two individual RMSDs:

$$\text{RMSD}_{\text{total}} = \text{RMSD}_{\text{RA-tox}} + \text{RMSD}_{\text{PS-RA}}. \tag{S4}$$

Fitting was performed by minimizing the $RMSD_{\text{total}}$. As shown in Figs. 11(a) and (b), the best-fit parameters were found to be $\lambda_A = 0.44$ nm, $\lambda_B = 0.1$ nm, and $DOS_B/DOS_B = 1500$, resulting in $RMSD_{\text{total}} = 1.73$. The corresponding fitting curves are shown as blue dashed lines in Figs. S7(a) and S7(b).

To evaluate the reasonable range of parameter variation, we defined acceptable fittings as those within a 20% increase of the minimum $RMSD_{\text{total}}$ value, while keeping $P_{S,A} = 32\%$ and $P_{S,B} = 0\%$ fixed. Within this criterion, the following parameter ranges were obtained: $\lambda_A = 0.33\sim0.47$ nm, $\lambda_B = 0.12 \sim 0.084$ nm, and $DOS_B/DOS_B = 200\sim10000$. The lower and upper bounds of these ranges correspond to the fittings with $RMSD_{\text{total}} = 2.11$ and 2.01, as the orange and pink dashed curves shown in Figs. S7(a) and S7(b), respectively.

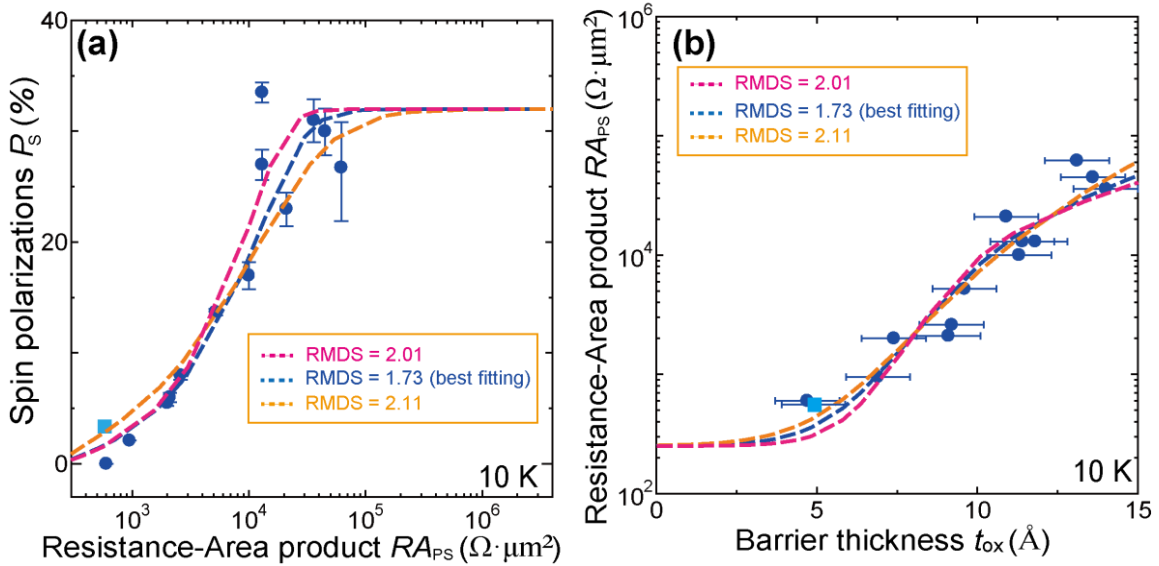

**FIG. S7.** Experimental (a) $P_S - RA_{PS}$ and (b) $RA_{PS} - t_{ox}$ plots, extracted from Figs. 9(d) and (c), respectively. The blue dashed curves represent the best-fit results calculated from Eqs. (3) and (4) with $RMSD_{\text{total}} = 1.73$. The orange and pink dashed curves represent the fittings with $RMSD_{\text{total}} = 2.11$ and 2.01, respectively.